\documentclass[a4paper, 11pt]{article}

\pdfoutput=1
\usepackage{amssymb}
\usepackage{epsfig, graphicx, graphics, multicol}
\usepackage{hyperref}
\usepackage{color}
\definecolor{rosso}{cmyk}{0,1,1,0.4}
\definecolor{rossos}{cmyk}{0,1,1,0.55}
\definecolor{rossoc}{cmyk}{0,1,1,0.2}
\definecolor{blu}{cmyk}{1,1,0,0.3}
\definecolor{blus}{cmyk}{1,1,0,0.6}
\definecolor{bluc}{cmyk}{1,1,0,0.1}
\definecolor{verde}{cmyk}{0.92,0,0.59,0.25}
\definecolor{verdec}{cmyk}{0.92,0,0.59,0.15}
\definecolor{verdes}{cmyk}{0.92,0,0.59,0.4}

\def\circa#1{\,\raise.3ex\hbox{$#1$\kern-.75em\lower1ex\hbox{$\sim$}}\,}

\newcommand{\de}{{\rm d}}

\makeatletter
\def\art{\@ifnextchar[{\eart}{\oart}}
\def\eart[#1]#2#3#4#5#6{{\rm #2}, {\em #3 \rm #4} {\rm (#6) #5 ({\em #1})}}
\def\hepart[#1]#2{{\rm #2, \em#1}}
\newcommand{\oart}[5]{{\rm #1}, {\em #2 \rm #3} {\rm (#5) #4}}

\newcommand{\beq}{\begin{equation}}
\newcommand{\eeq}{\end{equation}}
\newcommand{\bea}{\begin{eqnarray}}
\newcommand{\eea}{\end{eqnarray}}
\newcommand{\ba}{\begin{array}}
\newcommand{\ea}{\end{array}}
\newcommand{\bi}{\begin{itemize}}
\newcommand{\ei}{\end{itemize}}
\newcommand{\bn}{\begin{enumerate}}
\newcommand{\en}{\end{enumerate}}
\newcommand{\bc}{\begin{center}}
\newcommand{\ec}{\end{center}}

\newcommand{\nn}{\nonumber\\}

\newcommand{\gsim}{\lower.7ex\hbox{$\;\stackrel{\textstyle>}{\sim}\;$}}
\newcommand{\lsim}{\lower.7ex\hbox{$\;\stackrel{\textstyle<}{\sim}\;$}}
\newcommand{\cO}{\mathcal{O}}

\begin{document}

\tolerance=100000
\thispagestyle{empty}

\begin{flushright} CERN-PH-TH/2007-208\\ LA-UR-07-7323 \\ MIT-CTP-3889\end{flushright}
\vspace*{2.5cm}

\centerline{\Large \bf Quantum Resonant Leptogenesis 
and Minimal Lepton Flavour Violation}
\vspace*{1.5cm}

\centerline{Vincenzo Cirigliano $^{\rm a}$, Andrea De Simone $^{\rm b}$,  Gino Isidori $^{\rm c}$, 
Isabella Masina $^{\rm d}$ and Antonio Riotto $^{\rm d,e}$}

\vskip 1cm

\begin{flushleft}
\textit{$^{\rm a}$ Theoretical Division, Los Alamos National Laboratory, Los Alamos NM 87545, USA\\  
	$^{\rm b}$ Center for Theoretical Physics,
	Massachusetts Institute of Technology,\\ \hspace{0.2cm} Cambridge, MA 02139, USA \\
	$^{\rm c}$ Scuola Normale Superiore, Piazza dei Cavalieri 7, I-56100 Pisa, Italy, \\
         \hspace{0.2cm}  {\rm and} INFN, Laboratori Nazionali di Frascati, Via E. Fermi 40, I-00044 Frascati, Italy \\
        $^{\rm d}$ CERN, 
Department of Physics, Theory Division, CH-1211 Geneva 23, Switzerland \\ 
        $^{\rm e}$ INFN, Sezione di Padova,
via Marzolo 8, 35131 Padova, Italy }\\
\vskip 0.5cm
Electronic addresses: {\tt cirigliano@lanl.gov}, {\tt andreads@mit.edu}, {\tt Gino.Isidori@lnf.infn.it}, {\tt imasina@mail.cern.ch}, 
{\tt riotto@mail.cern.ch}
\end{flushleft}

\vskip 2cm
\begin{center}{\bf Abstract} \end{center}
\begin{quote}
It has been recently shown that the 
quantum Boltzmann equations may be relevant for the leptogenesis scenario. 
In particular, they lead to a time-dependent CP asymmetry which depends 
upon the previous dynamics of the system. This memory effect in the 
CP asymmetry is particularly important in resonant leptogenesis 
where the asymmetry is generated by the decays of nearly mass-degenerate 
right-handed neutrinos. We study the impact of the non-trivial time 
evolution of the CP asymmetry 
in the so-called Minimal Lepton Flavour Violation framework where
the 
charged-lepton and the neutrino Yukawa couplings are the 
only irreducible sources of lepton-flavour symmetry 
breaking  and resonant 
leptogenesis is achieved. 
We show that significant quantitative differences arise with 
respect to the case in which the 
time dependence of the CP asymmetry is neglected.
\end{quote}

\newpage
\setcounter{page}{1}

\section{Introduction}
\label{sect:introduction}

The observed baryon number asymmetry 
of the Universe $\eta_B\equiv n_B/n_\gamma =\left(6.3\pm 0.3\right)
\times 10^{-10}$ \cite{wmap} can be explained by 
the mechanism of thermal leptogenesis \cite{FY,leptogen,work}, 
the simplest implementation of this mechanism being realised by adding to the Standard Model (SM)
three heavy right-handed (RH) neutrinos, {\it i.e.}
in the framework of the seesaw \cite{seesaw}. In thermal leptogenesis  the heavy RH neutrinos
are produced by thermal scatterings after inflation  and subsequently
decay out-of-equilibrium in a lepton number and CP-violating way, 
thus satisfying Sakharov's constraints \cite{sakharov,baureview}. The 
dynamically generated  lepton asymmetry is 
then  converted into a baryon asymmetry
due to $(B+L)$-violating sphaleron interactions \cite{kuzmin}.

If  RH neutrinos are hierarchical in mass,
successful leptogenesis requires that 
the RH neutrinos are heavier than about  $10^9$ GeV \cite{di}.
Hence, the required reheating temperature after inflation cannot be much lower 
\cite{leptogen,davidsonetal1,aa,jmr}. In 
supersymmetric scenarios this may be in conflict with the upper bound
on the reheating temperature necessary to 
avoid the  overproduction of gravitinos during
reheating \cite{grav}. 
In the resonant leptogenesis scenario \cite{res} this tension may be
avoided. If the RH neutrinos are nearly degenerate in mass, self-energy 
contributions to the CP asymmetries may be  resonantly enhanced, thus
making thermal leptogenesis viable at  temperatures as low as the
TeV.  Resonant
leptogenesis seems therefore a  natural possibility. 
%%%%%%
%Nearly deg rh neutrinos are predicted in the M Lepton F V framework. The extend version.
%Leptogenesis has been studied in this framework -and the similar one of raditively induced resonant
%leptogenesis- 
%with flavour effects and without, also in connection with FCNC in rare lepton decays. 
%[The hypothesis of MFV consists in assuming that the Yukawa couplings are the only sources
% of breaking of the flavour symmetry.
%This has an unambiguos realization in the quark sector \cite{MQFV}. 
%In the lepton sector with heavy rh neutrinos (extended field content) 
%\cite{Cirigliano:2005ck, Cirigliano:2006su}the larger lepton-flavour symmetry group is
%$SU(3)\times SU(3) \times O(3)$. For Grand unification see \cite{Grinstein:2006cg}.
%Resonant leptogenesis as been considered in \cite{Cirigliano:2006nu} in the 1 flavour 
%approximation and in \cite{Branco:2006hz, Uhlig:2006xf} in the flavoured case.]
%%%%%%%%%%%%%%%%%%%%%

Nearly degenerate RH neutrinos naturally arise in 
the context of models satisfying 
the hypothesis of Minimal Flavour Violation (MFV)~\cite{MQFV,MFV,Cirigliano:2005ck}.
In the quark sector, where the MFV hypothesis has been formulated first, 
the MFV ansatz states that the two quark Yukawa couplings are the only 
irreducible breaking sources of the flavour-symmetry group defined by 
the gauge-kinetic Lagrangian~\cite{MFV}. In generic models satisfying this 
hypothesis, quark  Flavour Changing  Neutral Currents (FCNC) are naturally suppressed 
to a level comparable to experiments and new degrees of freedom can 
naturally appear in the TeV range. The extension of the MFV hypothesis
to the lepton sector (MLFV) has been formulated in Ref.~\cite{Cirigliano:2005ck} 
(and further studied in Refs.~\cite{Cirigliano:2006su,Grinstein:2006cg,Davidson:2006bd}), 
where a number of possible scenarios, depending on the field content 
of the theory, have been identified. The case more similar to the quark sector and 
more interesting from the point of view of leptogenesis is the 
so-called extended field content scenario of Ref.~\cite{Cirigliano:2005ck}.
Here the lepton field content is extended by three heavy RH neutrinos
with degenerate masses at the tree level. 
The largest lepton-flavour symmetry group 
of the gauge-kinetic term is $G_{\rm max} = SU(3)_{\ell } \times SU(3)_{e} 
\times O(3)_{N}$ and, according to the MLFV hypothesis, it is assumed that 
this group is broken only by the charged-lepton and neutrino 
Yukawa couplings $\lambda_e$ and $\lambda_\nu$. 
In relation to leptogenesis, the key feature of this scenario is that the degeneracy of 
the RH neutrinos is lifted {\em only} by 
corrections induced by the Yukawa couplings,
so that we end up with a highly constrained version of resonant leptogenesis.
Within this set up,  the viability of leptogenesis has  first been considered in the 
one-flavour approximation~\cite{Cirigliano:2006nu} and subsequently in the 
flavoured case~\cite{Branco:2006hz, Uhlig:2006xf}.

Resonant leptogenesis in models satisfying the MLFV hypothesis  is the subject of the present paper. 
Our  analysis turns out to be 
necessary 
in view of the recent results achieved by studying  
the dynamics 
of thermal leptogenesis by means of 
quantum Boltzmann equations \cite{dsr1} (for an earlier study, see
Ref. \cite{buchmuller}). Quantum Boltzmann equations were obtained 
starting from  the  non-equilibrium quantum field theory based on the 
Closed Time-Path   formulation and  have an obvious interpretation in terms of 
gain and loss processes.   However, they differ from the classical
Boltzmann equations since there appear integrals over 
the time where theta functions ensure that
the dynamics is causal. In  the classical kinetic theory the scattering terms 
do not include any integral over the past history of the 
system which is equivalent to assume that any collision in the plasma  
does not depend upon the previous ones. On the contrary,   
quantum distributions possess strong memory effects and the thermalization 
time obtained from the quantum transport theory may be 
substantially longer than the one obtained from 
the classical kinetic theory. 
Furthermore, and more importantly, the CP asymmetry turns out to be
 a function of time,  
its value at a given
instant depending  upon the previous history of the system. 
If the time variation of the CP asymmetry is shorter than the
relaxation time of the particles abundances, the solutions to the
quantum and the classical
Boltzmann equations are expected to 
differ only by terms of the order of the ratio
of the time-scale of the CP asymmetry to the relaxation time-scale of the
distribution. In thermal leptogenesis with hierarchical 
RH neutrinos this is typically the case. However, in the
resonant leptogenesis scenario, RH neutrinos  
 are almost degenerate
in mass and the CP asymmetry from the decay of the $i$-th RH neutrino $N_i$ 
is resonantly enhanced by the $j$-th neutrino if the mass difference
$(M_j-M_i)$
is of the order of the decay rate of the RH neutrinos. The typical time-scale
to build up coherently the time-dependent CP asymmetry is of the order of $(M_j-M_i)^{-1}$ 
\cite{dsr1,dsr2}, which
can be larger than the time-scale  
for the change of the abundance of the
RH neutrinos. This tells us that in the case of resonant leptogenesis
significant differences are expected between the classical and the
quantum approach. 

In this paper we perform a detailed study of the role of quantum memory  effects in the
resonant leptogenesis scenario within the MLFV hypothesis showing that
memory effects may substantially change the prediction for the baryon asymmetry. We consider both the 
normal hierarchical (NH) and the inverse hierarchical (IH) cases for light neutrinos
and also consider the role played by flavour in leptogenesis \cite{f1,f2,f3,f4}. 

Our analysis is organized as follows. In  Section 2 we provide a brief summary of
thermal leptogenesis, focussing on the impact of the quantum memory effects. Section 3
contains the key ingredients of the MLFV scenario. In Section 4 we set the stage for the
study of the 
relevance of non-equilibrium quantum effects
in MLFV-leptogenesis.  
We then proceed discussing
the impact of quantum effects on  the two scenarios
in which CP violation arises from the RH sector in   Section 5, 
and only from the PMNS matrix in  Section 6.
We give our conclusions in Section~\ref{sect:conclusions}.

\section{A brief summary of thermal leptogenesis}
\label{sect:mlfv} 

In order to set the stage for the subsequent analysis about 
the impact of non-equilibrium quantum effects in the MLFV framework, here we first 
briefly recall the general features of the thermal leptogenesis scenario, 
and later discuss the restrictions imposed by the hypothesis 
of Minimal Flavour Violation.

We consider a model where three right-handed (RH) 
neutrinos $N_{i}$ ($i=1,2,3$) with Majorana masses $M_{3} \ge M_{2} \ge M_{1}$
and Yukawa couplings ${\lambda_\nu}$ are added to the Standard Model (SM) Lagrangian.
Working in the basis in which the Yukawa couplings for the charged leptons $\lambda_e$ 
are diagonal, the Lagrangian density is given by 
\begin{equation}
\label{L}
{\cal L} = {\cal L}_{\rm SM} + \left(\frac{1}{2} M_{i} ~N_i^2 + 
N_i~{(\lambda_\nu)}_{i\alpha} ~\ell_\alpha ~ H + 
\bar{e}_{\alpha} ~{(\lambda_e)}_\alpha ~ \ell_\alpha ~H^c+\hbox{h.c.}\right)\, ~~,
\label{Lag}
\end{equation}
where $\ell_\alpha$ and $e_{\alpha}$ indicate the lepton doublet and singlet with flavour 
$(\alpha=e,\mu,\tau)$ respectively, and $H$ is the Higgs doublet whose
neutral component has a vacuum expectation value equal to $v=174$ GeV. 
For heavy RH neutrinos ($M_i \gg v$), light neutrino masses are  
obtained via the see-saw (type I) mechanism \cite{seesaw}
\begin{equation}
m_\nu = U^* \hat m_\nu  U^\dagger = \lambda_\nu^T \hat M_\nu^{-1} \lambda_\nu~ v^2~~,
\label{lamnu}
\end{equation}
where $U$ is the PMNS mixing matrix  (and we adopt the convention of placing a hat over diagonal 
matrices with real and positive elements).
In this framework  the baryon asymmetry is  generated by  
weak sphaleron  processes  converting
the non-zero lepton number  produced by out-of-equilibrium decays of the heavy RH
neutrinos.    

The key quantities controlling the production of a net lepton number 
are the CP violating  asymmetries in the $N_i$ decay rates

\beq
\epsilon_{i \alpha} \equiv \frac{
\Gamma (N_i \to \ell_\alpha \bar{H}) - \Gamma (N_i \to \bar{\ell}_\alpha H)
}{
\Gamma (N_i \to \ell_\alpha \bar{H}) + \Gamma (N_i \to \bar{\ell}_\alpha H)
}~.
\eeq
The inclusion of quantum effects discussed in the introduction (for technical details see Refs.~\cite{dsr1,dsr2})
introduces a time dependence in the CP asymmetry
\bea  
\epsilon_{i\alpha}(z)&=& \sum_{j\neq i} \epsilon^{(j,i)}_{\alpha}\,
 m^{(i,j)}(z), \nonumber \\ 
m^{(i,j)}(z)&=&2 ~\sin^2\left(\frac{1}{2} \frac{M_j-M_i}{2 H(M_1)} z^2\right)
- \frac{\Gamma_j}{M_j-M_i}~\sin\left(\frac{M_j-M_i}{2 H(M_1)}z^2\right),\nonumber\\
\epsilon^{(j,i)}_{\alpha}&=&\frac{1}{8\pi}
\frac{\textrm{Im}\left[ (\lambda_\nu)_{i\alpha} (\lambda_\nu)^\dagger_{\alpha j} 
(\lambda_\nu \lambda_\nu^{\dagger})_{ij} \right] }
{\left(\lambda_\nu \lambda_\nu^{\dagger}\right)_{ii}}~ (g_s^{(j,i)}+g_v^{(j,i)}), \label{mij}\\
g_s^{(j,i)}&=&\sqrt{\frac{x_j}{x_i}} ~\frac{1}{1-\frac{x_j}{x_i}} ~\frac{1}{1+\frac{\Gamma_j^2/M_i^2}{(1-x_j/x_i)^2}},\nonumber\\
g_v^{(j,i)}&=&\sqrt{\frac{x_j}{x_i}} ~\left( 1- (1+\frac{x_j}{x_i})~ \ln \frac{1+x_j/x_i}{x_j/x_i} \right),\nonumber
\eea
where $\Gamma_j\equiv \sum_\beta \Gamma (N_j \to \ell_\beta \bar{H}) =(\lambda_\nu\lambda_\nu^\dagger)_{jj} M_{j}/(8 \pi)$ 
 is the total decay rate of the $j$-th RH neutrino,
$H$ is the Hubble expansion rate, $z=M_1/T$, $T$ denotes the temperature, 
$g_s$ and $g_v$ are the self-energy and the vertex correction functions respectively  and $x_i= (M_i/M_1)^2$ 
(a short summary of the Boltzmann equations can be found in the Appendix A).
The combination of Yukawa couplings appearing in 
Eq.~(\ref{mij})  is quite constrained under the hypothesis of MLFV 
and, as we discuss below, 
the requirement of non-vanishing  $\epsilon_{i \alpha}$
leads to non-trivial constraints on the free parameters of this framework. 

In the quantum Boltzmann approach, the typical time-scale for the variation of the CP asymmetry is 
\beq
t=\frac{1}{2H(T)}=\frac{z^2}{2H(M_1)}\sim \frac{1}{M_j-M_i}=\frac{1}{\Delta M_{ji}}~~.
\eeq
As a consequence, quantum effects are expected to be sizable if 
$1/\Delta M_{ji}$ is larger than the time-scale   for changing the abundance, $1/\Gamma_i$. 
In other words, the oscillation frequency $\Delta M_{ji}$ has to be sufficiently smaller 
than $\Gamma_i$, so that the factors $m^{(i,j)}(z)$ do not effectively average to  one.  
Under these conditions,  the amplitude of the ``sin" term in $m^{(i,j)}(z)$ is also enhanced, 
which will turn out to be a crucial effect. If  $1/\Delta M_{ji}$ is smaller than the time-scale
for changing the particle abundances, then the CP asymmetry may be averaged over many scatterings
\cite{dsr1} 
and  it reduces to the classical value $\epsilon_{i\alpha}= \sum_{j\neq i} \epsilon^{(j,i)}_{\alpha}$.

The above discussion leads us to formulate a quantitative criterion for the importance of 
quantum effects, namely  $\Delta M_{ji}  \lsim \Gamma_i$.
This  criterion can be naturally satisfied  if RH neutrinos are nearly degenerate, as first 
pointed out in  \cite{dsr2} and therefore  
in models  based on MLFV.   
So the next task  is to  identify  the parameters controlling  the ratio $\Delta M_{ji} / \Gamma_i$
within the MLFV framework.

\section{The MLFV scenario}
The MLFV approach with extended field content~\cite{Cirigliano:2005ck}
is based on the assumption that the largest possible flavour symmetry group
of the lepton sector,
$SU(3)_\ell \times SU(3)_e \times O(3)_N$, 
is broken only by two irreducible sources:
the Yukawa couplings $\lambda_e^0
%=\hat m_e/v
$ and $\lambda_\nu^0$ transforming as $(\bar 3, 3, 1)$ and $(\bar 3,1,3)$ respectively.

In the limit of vanishing Yukawa couplings, the $O(3)_N$ symmetry is exact and RH neutrinos are degenerate 
at the common scale $M_\nu^0$. Switching on the Yukawa interactions $\lambda^0_\nu$ and $\lambda^0_e$, 
the mass degeneracy is removed as an effect of those combinations of Yukawa couplings transforming as $(1,1,6_s)$
\cite{Cirigliano:2006nu}:
\bea
M_\nu & =& M_\nu^0 ~\left[ ~1 ~+~ c^{(1)} ~(h_\nu^0+{h_\nu^0}^T) 
~+~ c_{1}^{(2)} ~( (h^{0}_\nu)^2+({h^0_\nu}^T)^2) \right. \nn
& &~~~~~~~~~~\left. +~ c_{2}^{(2)} ~h^0_\nu {h^0_\nu}^T ~+~ c_{3}^{(2)} ~{h^0_\nu}^T h^0_\nu ~
+~ c_{4}^{(2)} ~( h^0_e + {h^0_e}^T )~+~ ....~\right] ~~,
\label{mnu}
\eea
where $h_\nu^0=\lambda^0_\nu {\lambda^0_\nu}^\dagger$, 
$h_e^0=\lambda^0_\nu {\lambda^0_e}^\dagger \lambda^0_e {\lambda^0_\nu}^\dagger$
and one can choose\footnote{The charged lepton and Dirac-neutrino Yukawa couplings themselves receive
small corrections induced by $\lambda^0_{\nu,e}$; the effect of going in the basis where charged leptons are diagonal
is accounted for by a redefinition of the $c$'s in Eq.(\ref{mnu}).}
 $\lambda_e^0=\hat m_e/v$.
The coefficients $c$'s are arbitrary parameters smaller than ${\cal{O}}(1)$. 
If one interprets them as arising from radiative effects, it is natural to have 
$c^{(1)}\sim 1/(4\pi)^2$ and $c^{(2)}_i \sim (c^{(1)})^2$.
From the unitary matrix $\bar U$ diagonalising $M_\nu$, one obtains $\hat M_\nu$ and $\lambda_\nu$
of Eqs.(\ref{Lag}) and (\ref{lamnu}): 
\beq
\hat M_\nu=\bar U M_\nu \bar U^T~~~~~,~~~~~~~  \lambda_\nu= \bar U \lambda^0_\nu.
\eeq
Notice that, due to the smallness of the $c$'s, it is natural to expect $\bar U$ to be close to the unity 
matrix (or possibly a permutation matrix, given the ordering convention $M_1 \le M_2 \le M_3$).
%Then one calculate $m_\nu=\bar \lambda^{T} \hat M_\nu^{-1} \bar \lambda v^2 = U^* \hat m_\nu U^\dagger$,
%It is convenient to introduce \cite{Casas:2001sr}
%the complex orthogonal matrix $R=\hat M_\nu^{-1/2} \lambda_\nu U \hat m_\nu^{-1/2} v$.

One can see $\lambda^0_\nu$ as the neutrino Yukawa coupling associated to degenerate RH neutrinos.
Accordingly, a convenient parameterisation to be exploited in the following is \cite{Casas:2001sr}
\beq
\lambda^0_\nu= \frac{1}{v}~\sqrt{M_\nu^0} ~H~ \sqrt{\hat m_\nu^0}~ {U^0}^\dagger~~,
\label{la0nu}
\eeq
where $\hat m_\nu^0={\rm diag}(m_1^0,m_2^0,m_3^0)$, 
$U^0=R_{23}(\theta^0_{23})\Gamma_{\delta^0} R_{13}(\theta^0_{13})\Gamma_{\delta^0}^\dagger R_{12}(\theta^0_{12})
\rm{diag}(e^{i\frac{\alpha^0_1}{2}},e^{i\frac{\alpha^0_2}{2}},1)$,
$\Gamma_{\delta^0}={\rm diag}(1,1,e^{i\delta^0})$ and $H$ is an orthogonal hermitian matrix 
\beq
H= e^{i\Phi}~~~~~,~~~~%=1-\frac{\cosh r-1}{r^2} \Phi^2 + i \frac{\sinh r}{r} \Phi~~~,~~~~
\Phi= \left( \matrix{ 0 & \phi_1 & \phi_2 \cr -\phi_1 & 0 & \phi_3 \cr -\phi_2 & -\phi_3 & 0} \right)~~~~.
%~~~~r= \sqrt{\sum_i \phi_i^2}~~.
\eeq
Clearly, the smaller  the splittings  among the RH neutrinos, the closer  $\hat m_\nu$ and $U$ 
are to  $\hat m_\nu^0$ and $U^0$ respectively. Since the approximations $\hat m_\nu \approx \hat m_\nu^0$ and
$U\approx U^0$ will turn out to be good, in all the subsequent discussion
we drop the $0$-superscript above the parameters of $\hat m_\nu^0$ and $U^0$.  
For later convenience, notice also that 
\bea
h^0_\nu &=&\lambda^0_\nu {\lambda^0_\nu}^\dagger= H \hat m^0_\nu H ~\frac{M_\nu^0}{v^2}~~,\\
h_e^0&=&\lambda^0_\nu {\lambda^0_e}^\dagger \lambda^0_e {\lambda^0_\nu}^\dagger = H E H ~\frac{M_\nu^0}{v^2}~~~,~~~
E=\sqrt{\hat m_\nu^0} {U^0}^\dagger \left( \frac{\hat m_e}{v} \right)^2 U^0 \sqrt{\hat m_\nu^0}~~.
\eea
%For normal hierarchy (NH) 
%it is convenient to vary $m_1^0$ in its allowed range with 
%$m_3^0=\sqrt{{m_1^0}^2+\Delta m^2_{21}+\Delta m^2_{32}}$, $m_2^0=\sqrt{{m_1^0}^2+\Delta m^2_{21}}$, while
%for inverted hierarchy (IH) it is convenient to vary $m_3^0$ with 
%$m_1^0=\sqrt{{m_3^0}^2-\Delta m^2_{21}+\Delta m^2_{23}}$, $m_2^0=\sqrt{{m_3^0}^2+\Delta m^2_{23}}$.

\section{General Implications of  MLFV for thermal leptogenesis}
The MLFV scenario described above has six independent CP-violating phases.
As shown in Appendix ~\ref{appB} they can 
be characterized in terms of weak-basis invariants that in turn are 
in direct correspondence with linearly independent combinations of the asymmetries 
$\epsilon_{i \alpha}$. 
The analysis of weak-basis invariants (see Appendix~\ref{appB} for more details)
allows us to draw the following general conclusions on the viability of leptogenesis 
both in the one-flavour regime ($M_\nu^0 > 10^{12}$ GeV)  as well 
as in the flavour regime ($M_\nu^0 < 10^{12}$ GeV): 
\begin{itemize}
\item In the one-flavour regime, in order to produce the  lepton asymmetry 
one needs non-zero second-order RH  mass splitting ({\it i.e.} $c^{(2)}_i \neq 0$)
and $H \neq I$, {\it i.e.} CP violation (CPV)  in the RH sector
~\cite{Cirigliano:2006nu}.

\item  Flavour effects open at least in principle two new regimes for 
MLFV-leptogenesis which are not allowed in the one-flavour limit:  
(i) the case in which RH mass splitting is induced only (or mainly) by $c^{(1)}$. 
This situation requires $H\neq I$, 
namely CPV in the RH sector;
(ii) the case in which CPV arises {\em only} from PMNS phases~\cite{Uhlig:2006xf},
 namely $H=I$ and $M_\nu$ is real. 
In this case  the lepton asymmetry is proportional to $c^{(2)}_4$.         
%  vanishes unless $c^{(2)}_4 \neq 0$.         
\end{itemize}
We first study the Yukawa-induced RH mass splitting. 
The largest correction term  
is naturally expected to be the one associated to $c^{(1)}$ in Eq.~(\ref{mnu}), so that many 
considerations about the RH neutrino spectrum and the frequency of the oscillations in $m^{(i,j)}$ 
can already be drawn. 
This program is straightforward in the limit of small $\phi_i$, 
which from the numerical analysis will turn out to be the relevant one in order to 
achieve  viable leptogenesis. 
%%%%%%%%%%%%%
%  because, as will turn out that from the numerical analysis,
 %  small $\phi_i$ are needed 
% for leptogenesis to be viable. 
%%%%%%%%%%%%%%
In this regime one can then expand the relevant matrices in series of $\phi_i$. 
In particular, one has 
from Eq.(\ref{la0nu})
\beq
\lambda^0_\nu =   \frac{\sqrt{M_\nu^0}}{v}\left( \matrix{ 
e^{-i\frac{\alpha_1}{2}}{\sqrt{m_1}} c_{12}& -e^{-i\frac{\alpha_1}{2}}{\sqrt{m_1}}c_{23}s_{12}& e^{-i\frac{\alpha_1}{2}}{\sqrt{m_1}}s_{12}s_{23}  \cr
e^{-i\frac{\alpha_2}{2}}{\sqrt{m_2}} s_{12}& e^{-i\frac{\alpha_2}{2}}{\sqrt{m_2}}c_{12}c_{23}& -e^{-i\frac{\alpha_2}{2}}{\sqrt{m_2}}c_{12}s_{23} \cr 
                          \sqrt{m_3} s_{13} e^{i\delta} & \sqrt{m_3} s_{23} & \sqrt{m_3} c_{23} }  \right) + {\cal O}(\phi_i),
\eeq
where the $0$-superscript is understood for $m_i$ and the parameters of $U^0$.  
In addition
\beq
h^0_\nu = \frac{M_\nu^0}{v^2}  \left( \matrix{ 
                   m_1 & i \phi_1 (m_1+m_2) & i\phi_2 (m_1+m_3)  \cr
                   -i\phi_1(m_1+m_2) & m_2 & i\phi_3 (m_2+m_3) \cr 
                   -i\phi_2 (m_1+m_3) & -i\phi_3 (m_2+m_3) & m_3 }  \right) + {\cal O}(\phi_i \phi_j)
\label{hzero}                   
\eeq
and
\bea
h^0_\nu+{h^0_\nu}^T&=& \frac{M_\nu^0}{v^2}\times\nonumber\\
&&\times  \left( \matrix{ 
                  2 m_1 & \phi_2\phi_3 (m_1+m_2+2m_3) & -\phi_1\phi_3 (m_1+2 m_2+m_3)  \cr
                  \phi_2\phi_3(m_1+m_2+2m_3) & 2m_2 & \phi_1\phi_2 (2m_1+m_2+m_3) \cr 
                  -\phi_1\phi_3 (m_1+2m_2+m_3) & \phi_1\phi_2 (2m_1+m_2+m_3) & 2 m_3 }  \right)\nonumber \\
&&+ \,{\cal O}(\phi_i^3).
\eea
The leading correction for RH neutrino masses is simply linked to the light neutrino mass spectrum
\beq
M_\nu=M_\nu^0 ~\left[ 1+ 2 c^{(1)}    \frac{ M_\nu^0 \hat m_\nu^0}{v^2} 
\left(1 + {\cal O}(\phi_i\phi_j) \right) \right].
\label{Mnu}
\eeq
Let us consider NH and IH for light neutrinos in turn.
From the form of $M_\nu$ it turns out that in first approximation, 
\beq 
\bar U^{NH}\approx I~~~,~~~~~\bar U^{IH}\approx \left( \matrix{0&0&1\cr 1&0&0 \cr 0&1&0} \right) = \tilde{I}~~.
\label{zz}
\eeq
Aiming at a compact notation, we introduce the matrix ${\bf m}={\rm diag}(m_l, m_i, m_h)$, where $m_l<m_i<m_h$ 
(light, intermediate, heavy); clearly, for NH $l=1,i=2,h=3$ while for IH $l=3,i=1,h=2$.
Then, defining $\Delta {\bf m}_{ji}={\bf m}_j-{\bf m}_i$ one has from Eq.(\ref{Mnu}):
\beq
M_i \approx M_\nu^0 \left(1 + 2 c^{(1)}~ \frac{M_\nu^0 {\bf m}_i}{v^2}\right)~~~,~~~~
\Delta M_{ji}\approx  2 c^{(1)} \Delta {\bf m}_{ji}~\left(\frac{M_\nu^0 }{v}\right)^2~~~~~,
\eeq
So we have identified the parameters that control 
the frequency of the oscillating $m^{(i,j)}$, see Eq.(\ref{mij}).
Moreover, 
the RH neutrino widths  and the  amplitude $A_{ji}$  of the sin-term in $m^{(i,j)}$
can then be expressed as 
\beq
\Gamma_j \approx \frac{ {\bf m}_j}{8\pi}\left(\frac{M_\nu^0}{v}\right)^2~~~~,~~~~~~
A_{ji} \equiv \frac{\Gamma_j}{\Delta M_{ji}}\approx \frac{1}{16\pi~c^{(1)}} \frac{{\bf m}_j}{ \Delta {\bf m}_{ji}} ~~.
\eeq

We are now in a position to study the condition  for the importance of quantum effects in the
CP asymmetries, 
namely $\Delta M_{ji}< \Gamma_i$.
Using the expressions above,  in the MLFV framework this condition can be cast into 
\beq
16\pi~c^{(1)} \frac{ \Delta {\bf m}_{ji}}{{\bf m}_i}<1~~,
\eeq
implying that  the difference between the classical and quantum approach will
increase as  $c^{(1)}$ becomes smaller and/or the pair ${\bf m}_j$-${\bf m}_i$ 
becomes degenerate.   
In the next sections we will study numerically and analytically the dependence of the baryon
asymmetry on $c^{(1)}$ and the lightest neutrino mass $m_l$, controlling the 
degeneracy of the light neutrino spectrum.

Both in the classical and quantum case,
in order for $\epsilon_\alpha^{(j,i)}$ to be non vanishing, it is crucial to have 
non diagonal entries in $h_\nu=\lambda_\nu \lambda_\nu^\dagger =\bar U h_\nu^0 \bar U^\dagger$. 
Since $\bar U$ is a small rotation, one has $h_\nu \sim h_\nu^0$.
From Eq.(\ref{hzero}) it turns out that the simplest situation in which 
$h_\nu$ is non-diagonal occurs when at least one among the $\phi_i$ is non vanishing, namely $H\neq 1$.
On the other hand, in the case $H= 1$, small  {\it real} non diagonal elements for $h_\nu$ have to be generated as an effect of $\bar U$;
in this case the CP violation needed for a non-zero asymmetry  arises 
from the  PMNS phases in $(\lambda_{\nu})_{i \alpha}$.
These two situations  are going to be discussed separately in the next sections.

Before turning to analyse the differences between the quantum and classical approaches 
focussing on these two relevant cases, we show the expressions for the washout
parameters. For both NH and IH, the washout parameters  $K_\alpha=\sum_i K_{i\alpha}$ and $K_i=\sum_\alpha K_{i\alpha}$
(see appendix \ref{Boltz} for more details) are given by: 
\bea
K_e &=&(m_1 c_{12}^2+m_2 s_{12}^2 +m_3 s_{13}^2)/m_*~~~,~~~~
\label{Ke} \\
K_\mu&=&(m_1 c_{23}^2 s_{12}^2+m_2c_{12}^2 c_{23}^2+m_3 s_{23}^2)/m_*~~~,~~~
\label{Km} \\
K_\tau&=&(m_1 s_{12}^2 s_{23}^2+m_2c_{12}^2 s_{23}^2 +m_3 c_{23}^2)/m_*~~~,
\label{Kt} \\
K_i&=&{\bf m}_i /m_* ~~~(i=1,2,3)~~~~. 
\label{Ki}
\eea
It is clear that none of the $K_\alpha$ $(\alpha=e,\mu,\tau)$ can be small:
the weak-washout regime cannot be obtained in the framework of MLFV. However,
let us emphasise that for NH the potentially smallest washout parameter is $K_e$. 
On the contrary, for IH all the $K_\alpha$ are comparable and relatively large 
(in particular, for small $m_3$ one has the relation $K_e \sim m_1/m_*\sim 2 K_{\mu,\tau}$).

In what follows, we  shall focus our attention on  the full flavoured regime\footnote{For
$M_\nu^0 > 10^{9}$ GeV the results do not change significantly.},  
$M_\nu^0  \lesssim 10^{9}$ GeV, 
and we shall illustrate the impact of quantum effects  
in the two   cases in which CP asymmetries arise                                        
 either from the RH ($H\neq I$) sector  or  
from the PMNS phases ($H=I$).

%%%%%%%%%%%%%%%%%%%%%%%%%%%%%%%%%%%%%%%%%%%%%%%%%%%%%%%%%%%%%%%%%%%%%%%%%%%%%%%%%%%%%%%%%%%%

\section{CP Violation  from $H \neq I$} 
\label{sect:Hneq1}

We first analyse the situation $c^{(1)}\gg c^{(2)}_i$ 
and $H\neq I$, in which the lepton asymmetries are 
proportional to the angles $\phi_i$. 
%As for the washout parameters one has, explicitly
%\bea
%K^{IH}_{2e}=K^{NH}_{1e}&=&m_1 c_{12}^2/ m_*~~,~~K^{IH}_{3e}=K^{NH}_{2e}=m_2 s_{12}^2 /m_*~~,~~
%K^{IH}_{1e}=K^{NH}_{3e}=m_3 s_{13}^2 /m_*~~,\no \\
%K^{IH}_{2\mu}=K^{NH}_{1\mu}&=&m_1 c_{23}^2 s_{12}^2 /m_*~~,~~K^{IH}_{3\mu}=K^{NH}_{2\mu}=m_2c_{12}^2 c_{23}^2 /m_*~~,
%~~K^{IH}_{1\mu}=K^{NH}_{3\mu}=m_3 s_{23}^2 /m_*~~,\no \\
%K^{IH}_{2\tau}=K^{NH}_{1\tau}&=&m_1 s_{12}^2 s_{23}^2 /m_*~~,~~K^{IH}_{3\tau}=K^{NH}_{2\tau}=m_2c_{12}^2 s_{23}^2 /m_*~~,~~
%K^{IH}_{1\tau}=K^{NH}_{3\tau}=m_3 c_{23}^2 /m_*~~,\no 
%\eea
For NH, let us focus on the asymmetries associated to the least washed-out flavour, 
that is the $e$-flavour. 
%$\epsilon_{ie}^{cl}$ for NH, 
Introducing the notation $F^{(j,i)}=1/(8\pi) (g_s^{(j,i)}+g_v^{(j,i)}) m^{(i,j)}$ and by working at first 
order in $\phi_i$, we find:
\bea
\epsilon_{1e} = \sqrt{\frac{m_2}{m_1}} \frac{m_1+m_2}{\tilde m} c_{12}s_{12} \phi_1 F^{(2,1)} \cos\left(\frac{\alpha_2-\alpha_1}{2}
\right)
              + \sqrt{\frac{m_3}{m_1}} \frac{m_1+m_3}{\tilde m} c_{12}s_{13} \phi_2 F^{(3,1)} \cos\left(\frac{2 \delta +\alpha_1}{2}\right),
\nonumber \\
\epsilon_{2e} = -\sqrt{\frac{m_1}{m_2}} \frac{m_1+m_2}{\tilde m} c_{12}s_{12} \phi_1 F^{(1,2)} \cos\left(\frac{\alpha_2-\alpha_1}{2}\right)
               + \sqrt{\frac{m_3}{m_2}} \frac{m_2+m_3}{\tilde m} s_{12}s_{13} \phi_3 F^{(3,2)} \cos\left(\frac{2 \delta +\alpha_2}{2}\right),
\nonumber \\
\epsilon_{3e} = -\sqrt{\frac{m_1}{m_3}} \frac{m_1+m_3}{\tilde m} c_{12}s_{13} \phi_2 F^{(1,3)} \cos\left(\frac{2\delta+\alpha_1}{2}\right)
                -\sqrt{\frac{m_2}{m_3}} \frac{m_2+m_3}{\tilde m} s_{12}s_{13} \phi_3 F^{(2,3)} \cos\left(\frac{2 \delta+\alpha_2}{2}\right),
\nonumber \\ \eea
where $\tilde m=v^2/M_\nu^0$. For zero CPV in PMNS the $\epsilon_{i\alpha}$ are maximised. 
Notice that for some values of the PMNS phases, one can suppress $\eta_B$;
however, if all $\phi_i\neq 0$, 
there is no choice of PMNS phases that allows all three $\epsilon$ to vanish\footnote{Indeed, 
the term proportional to $\{\phi_1,\phi_2,\phi_3\}$ vanishes when 
$\{\alpha_2-\alpha_1,2\delta+\alpha_1,2\delta+\alpha_2\}=\pi + {\rm mod}(2\pi)$, respectively.
The latter two conditions imply $\alpha_2-\alpha_1=0+{\rm mod}(2\pi)$, in conflict with the first condition.}.

\subsection{Dependences on $c^{(1)}$ and $m_l$}

\begin{figure}[t!]
\includegraphics[scale=0.5]{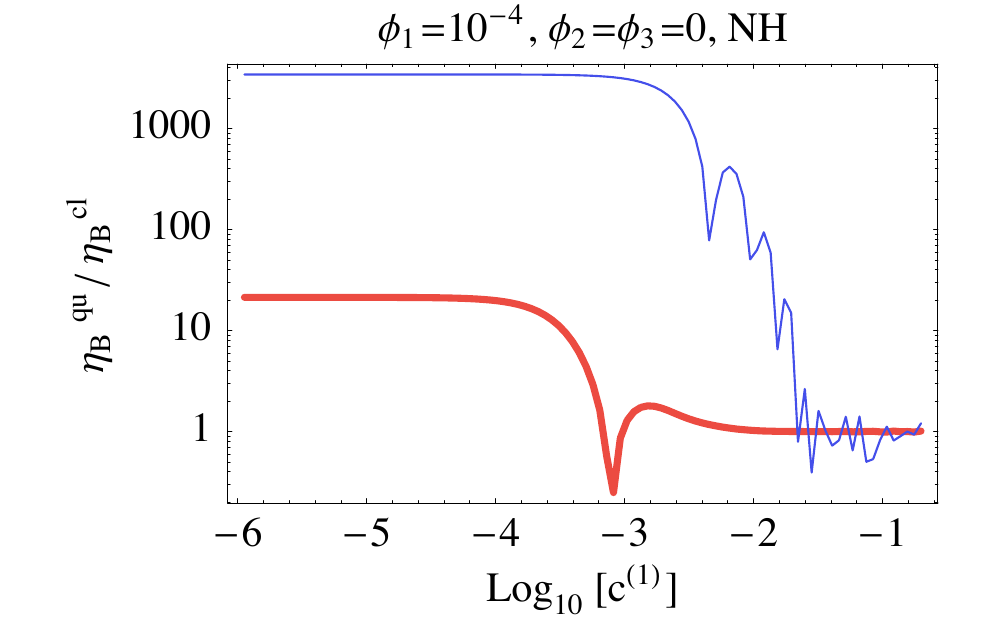}
\includegraphics[scale=0.5]{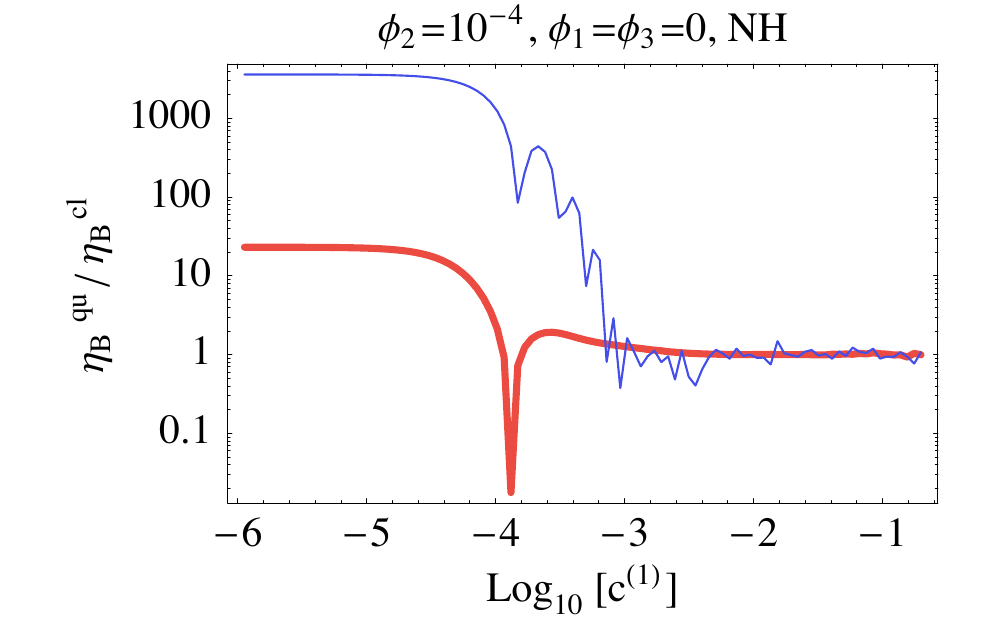}
\includegraphics[scale=0.5]{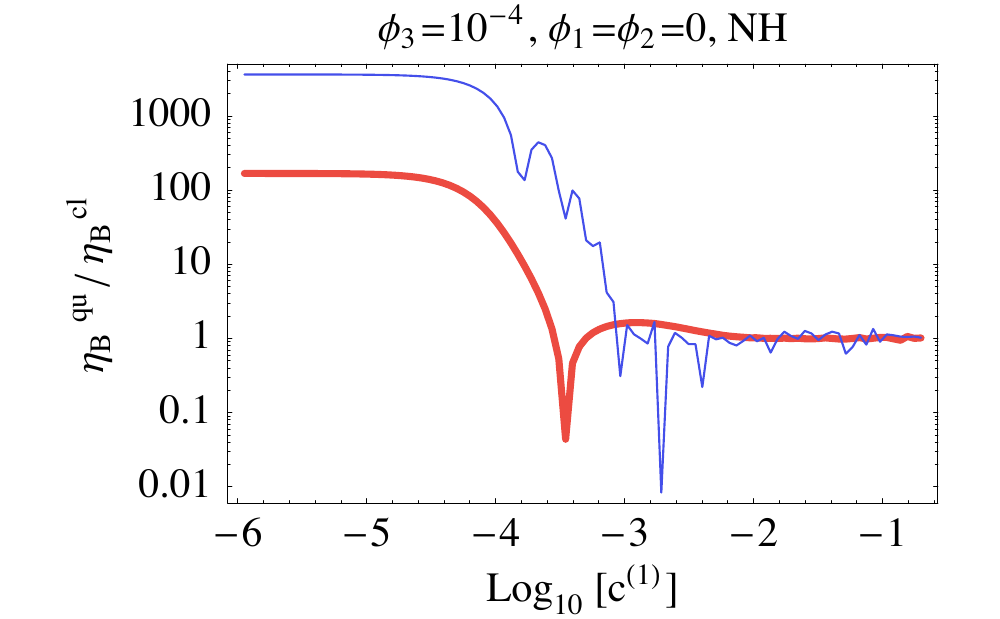}\\
\vskip5pt
\includegraphics[scale=0.5]{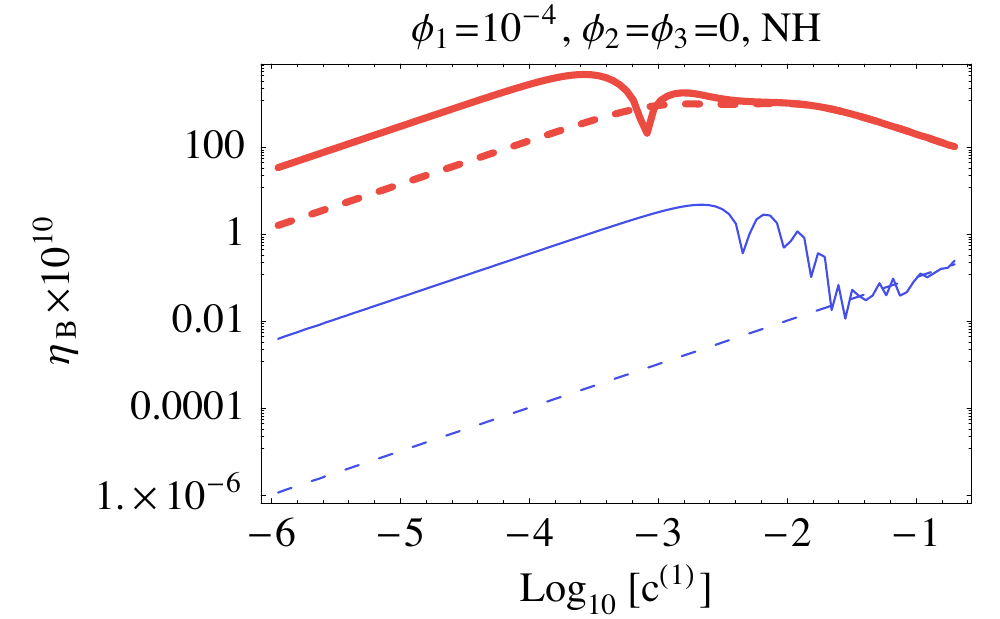}
\includegraphics[scale=0.5]{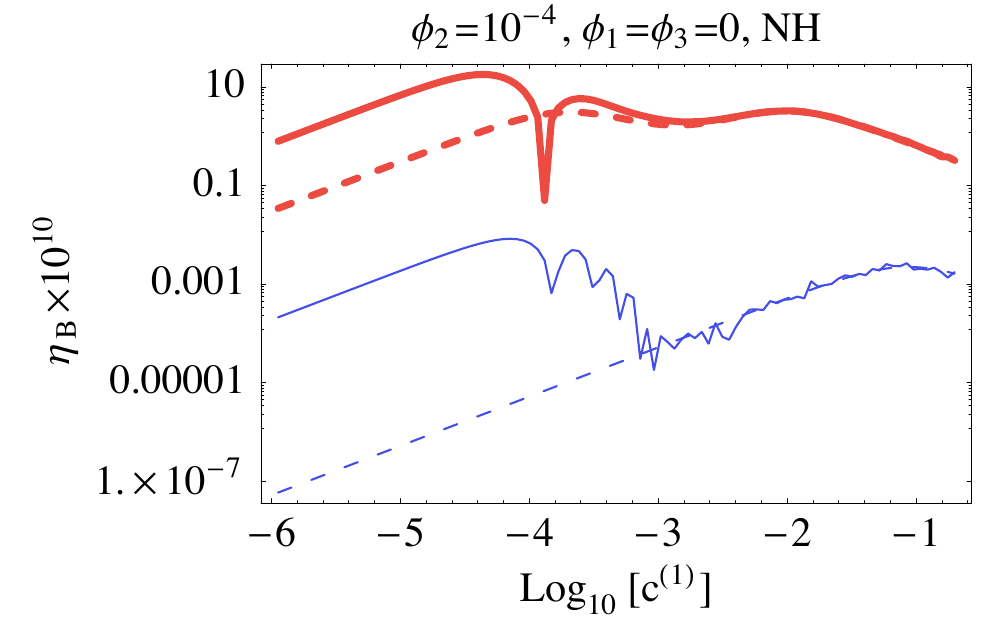}
\includegraphics[scale=0.5]{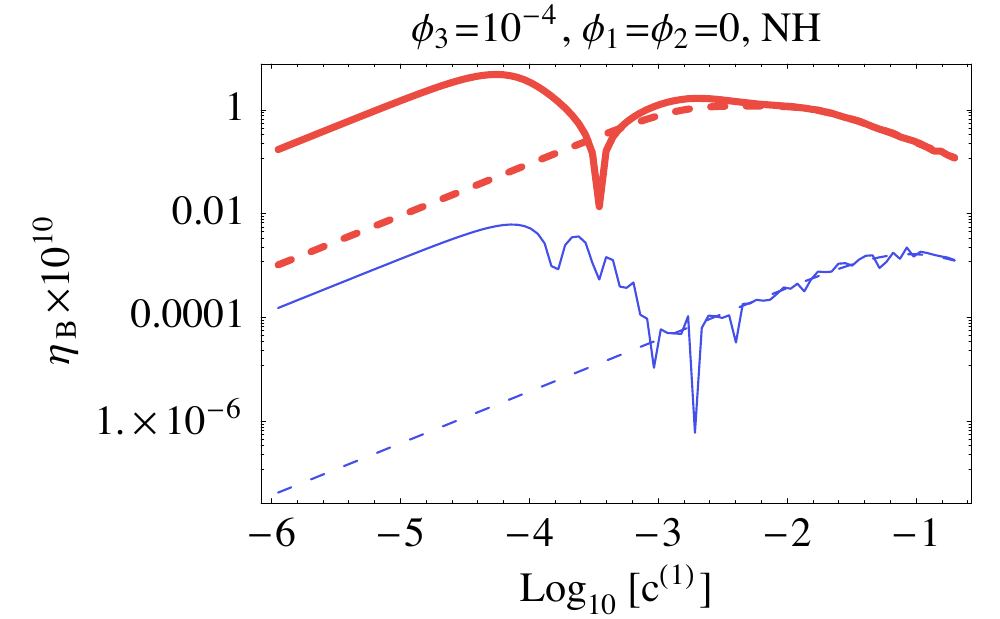}\\
\caption{ Dependence on $c^{(1)}$, in normal hierarchy. Thick red and thin blue lines  correspond to $m_1=10^{-3}$ eV and $m_1=10^{-1}$ eV, respectively.
{\it Top:} The absolute value of the ratio of the  baryon asymmetry with quantum effects ($\eta_B^{\rm qu}$) and without quantum effects ($\eta_B^{\rm cl}$).
{\it Bottom: } The absolute values of $\eta_B^{\rm qu}$  (solid lines) and $\eta_B^{\rm cl}$ (dashed lines).}\label{fig-NH-c1}
\end{figure}

\begin{figure}[h!]
\vskip 2cm
\includegraphics[scale=0.5]{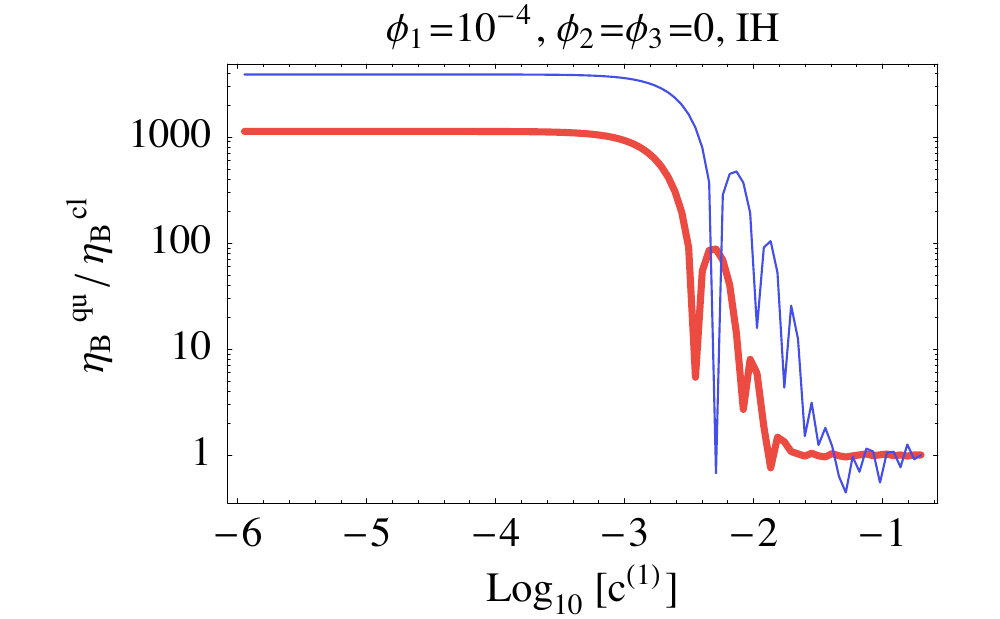}
\includegraphics[scale=0.5]{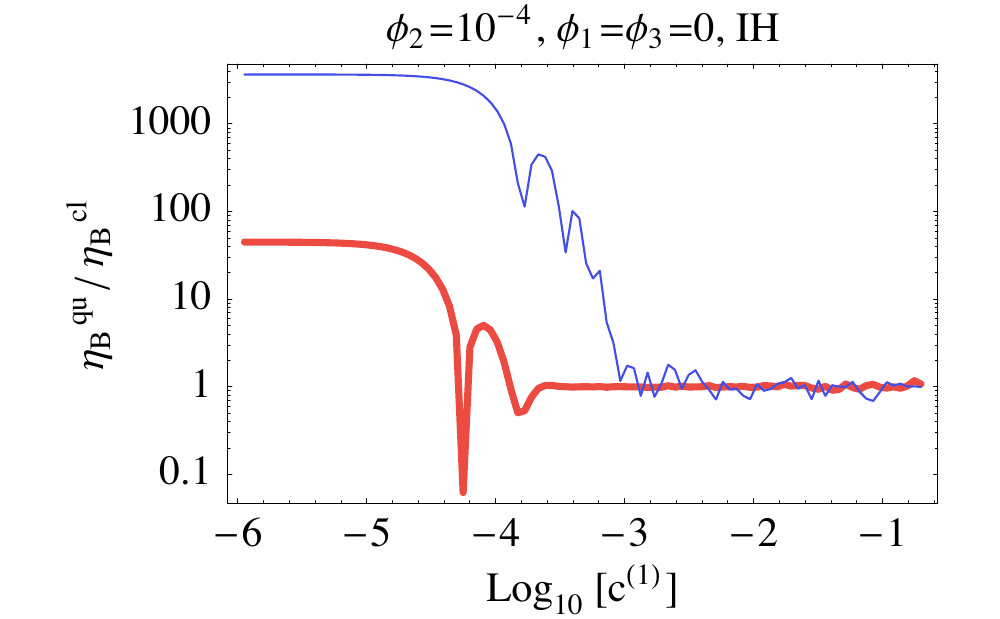}
\includegraphics[scale=0.5]{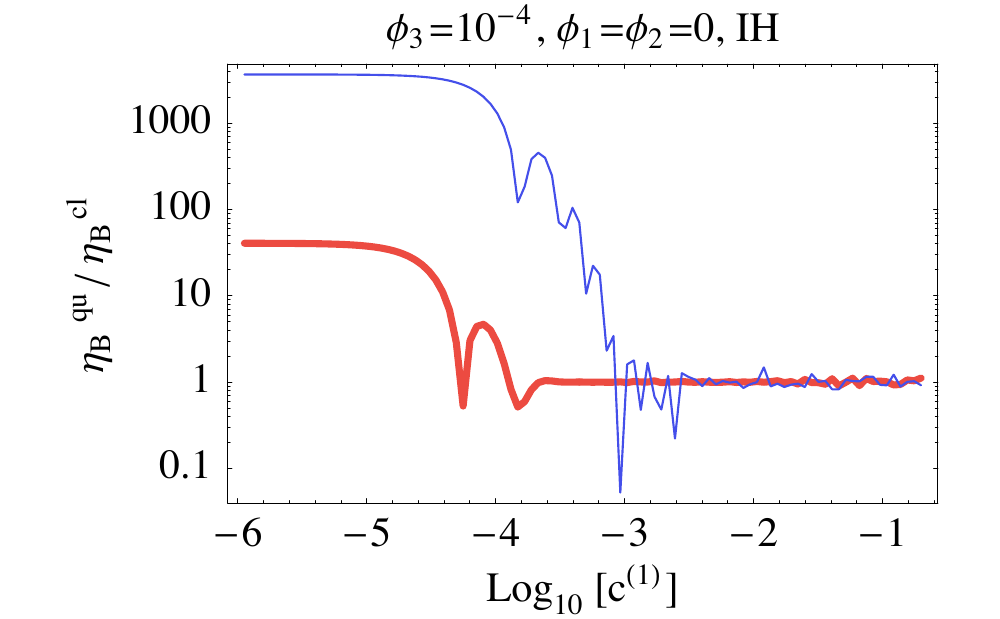}\\
\vskip5pt
\includegraphics[scale=0.5]{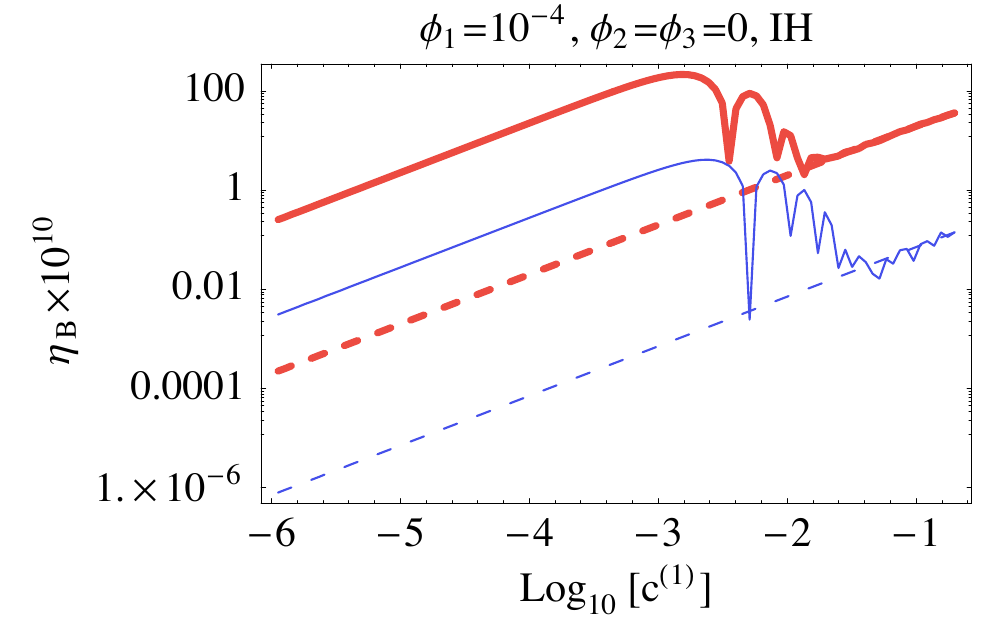}
\includegraphics[scale=0.5]{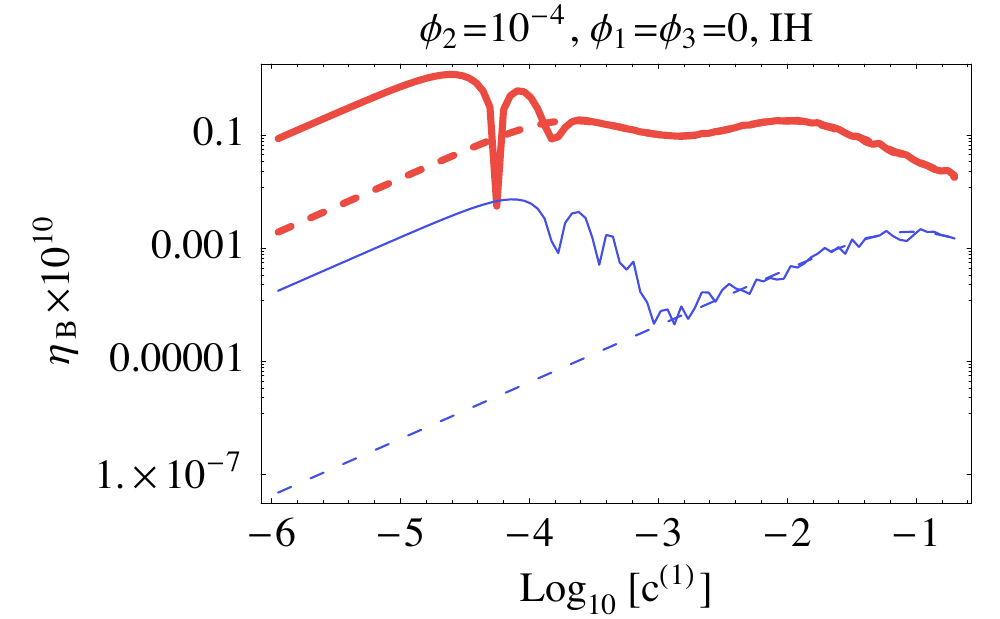}
\includegraphics[scale=0.5]{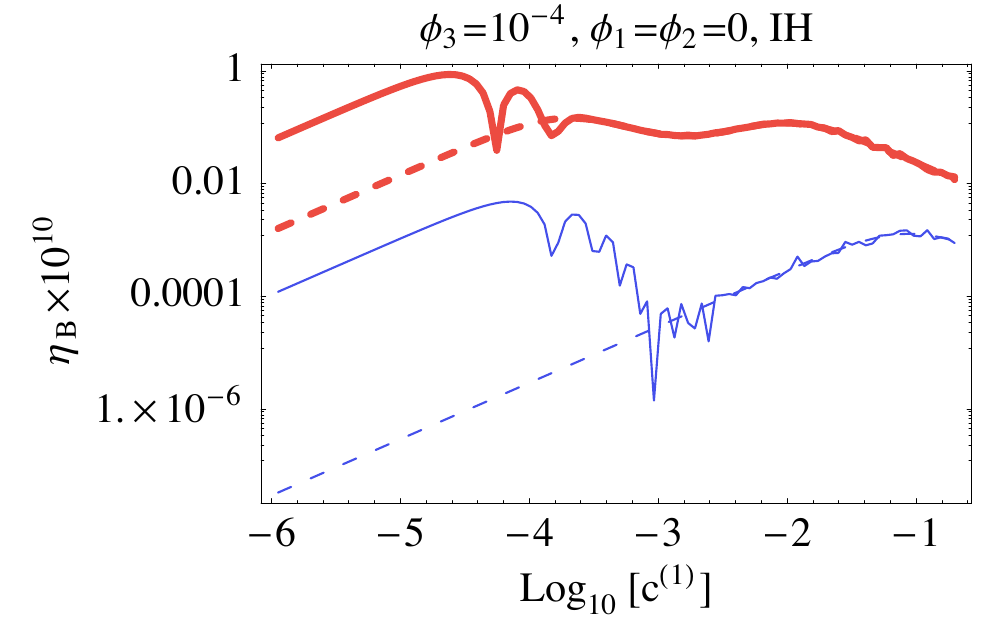}\\
\caption{Dependence on $c^{(1)}$, in inverse hierarchy. Thick red and thin blue lines  correspond to $m_1=10^{-3}$ eV and $m_1=10^{-1}$ eV, respectively.
{\it Top:} The absolute value of the ratio of the  baryon asymmetry with quantum effects ($\eta_B^{\rm qu}$) and without quantum effects ($\eta_B^{\rm cl}$).
{\it Bottom: } The absolute values of $\eta_B^{\rm qu}$  (solid lines) and $\eta_B^{\rm cl}$ (dashed lines).}\label{fig-IH-c1}
\end{figure}

\begin{figure}[t!]
\includegraphics[scale=0.5]{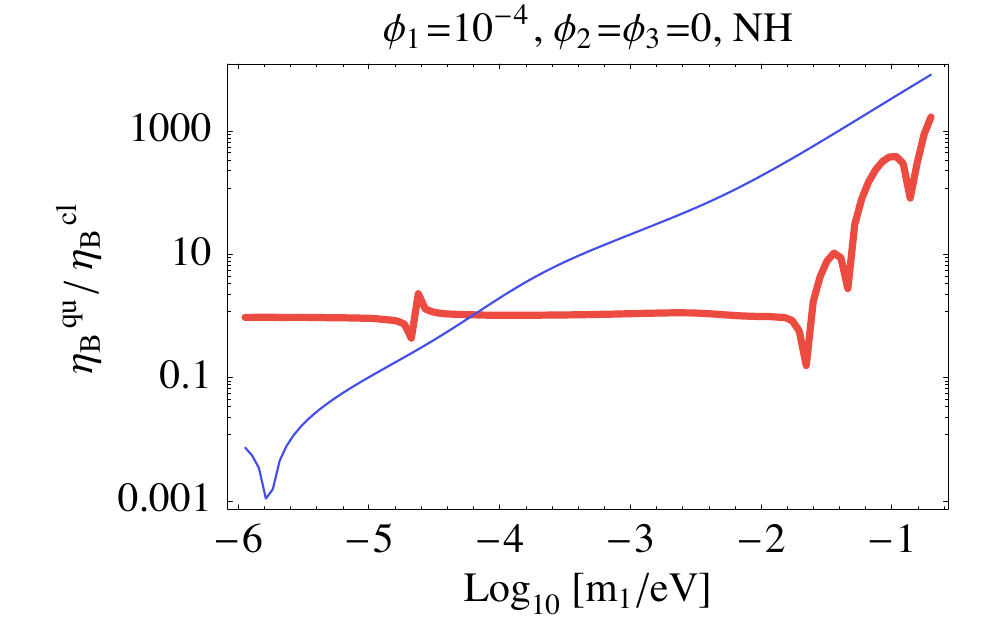}
\includegraphics[scale=0.5]{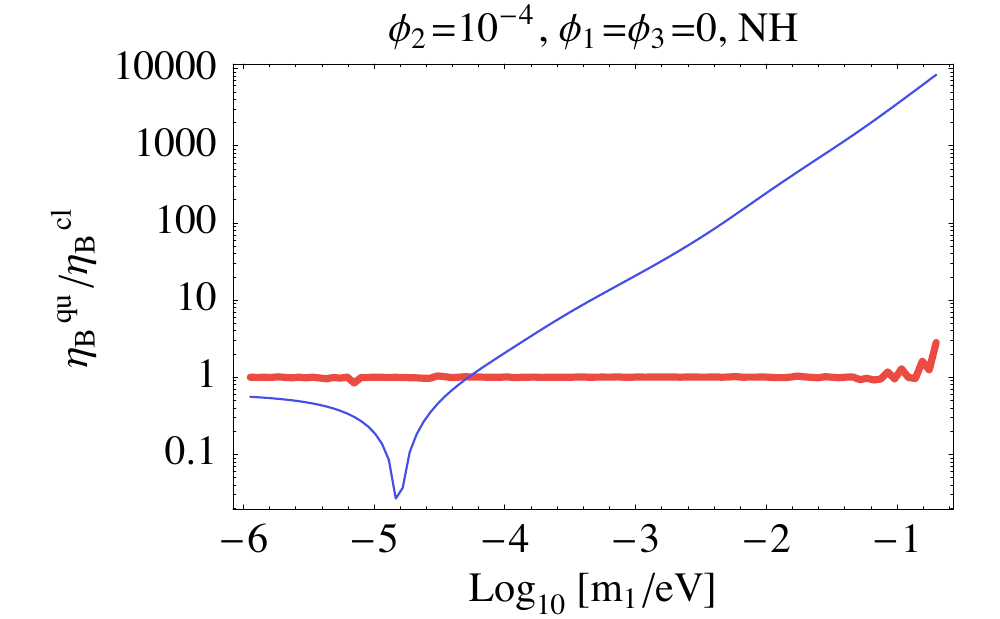}
\includegraphics[scale=0.5]{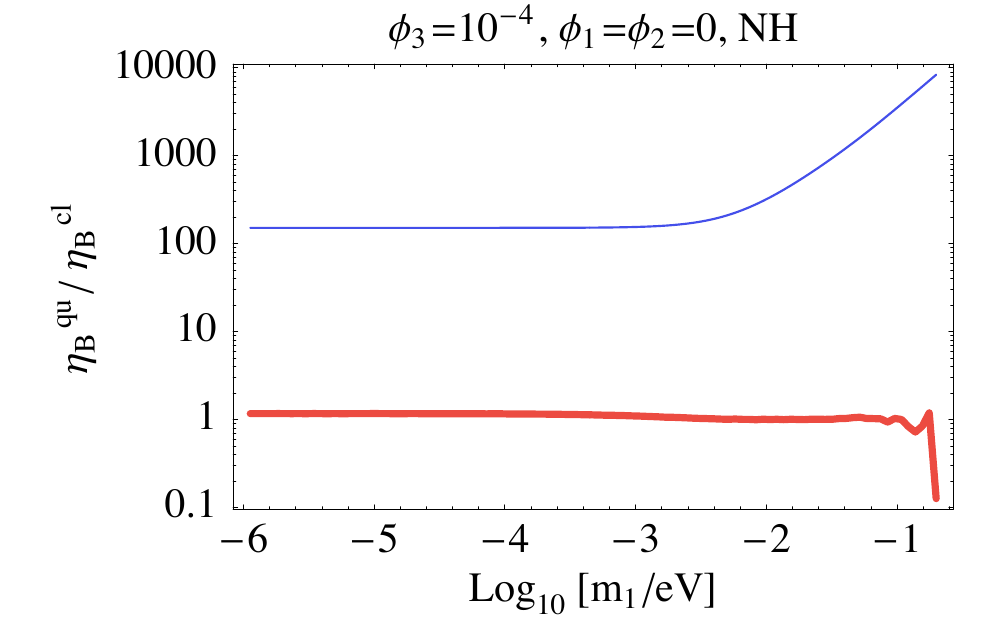}\\
\vskip5pt
\includegraphics[scale=0.5]{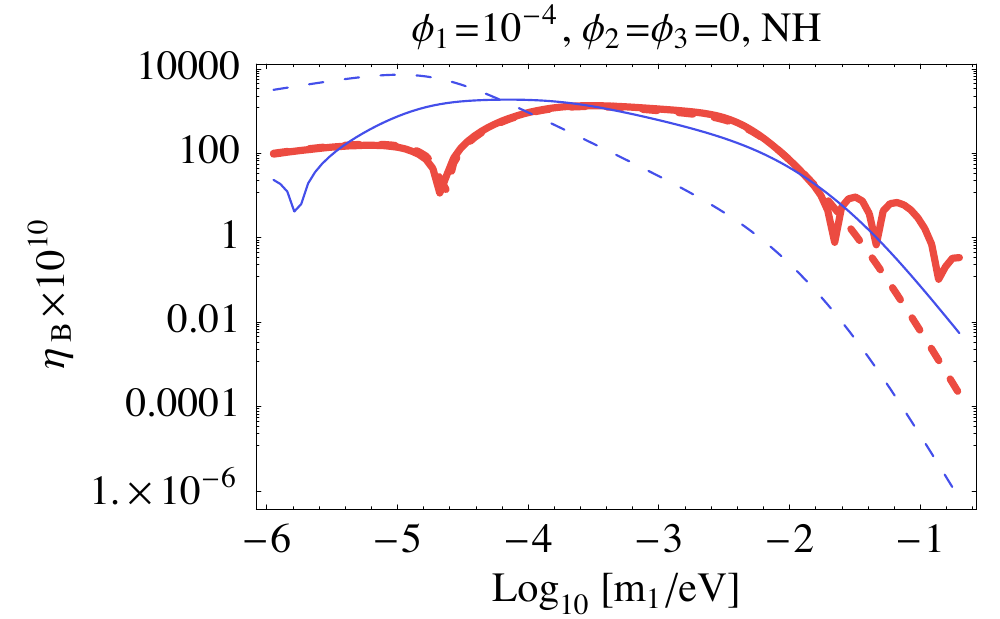}
\includegraphics[scale=0.5]{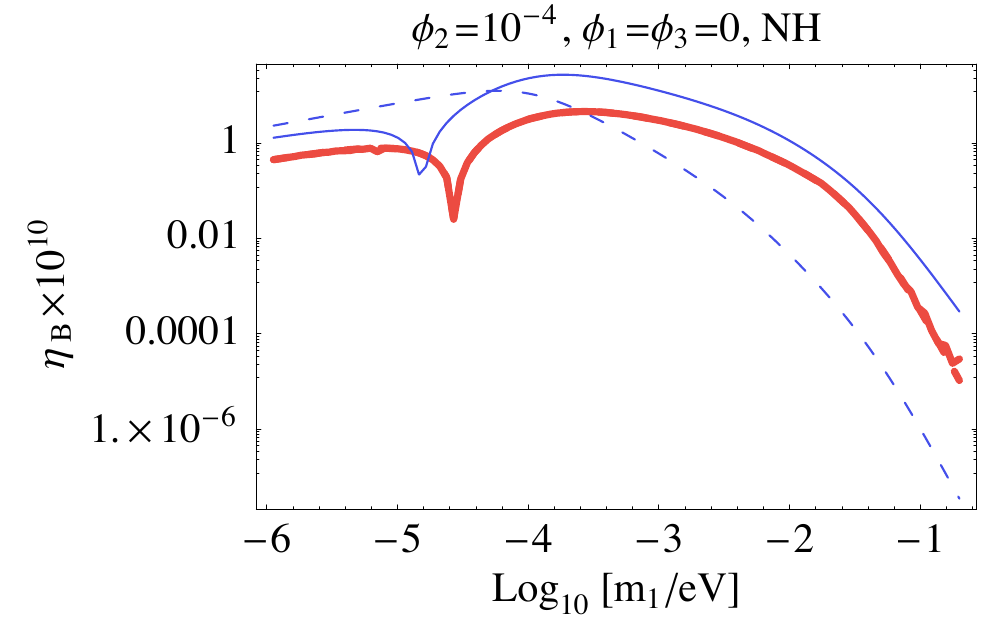}
\includegraphics[scale=0.5]{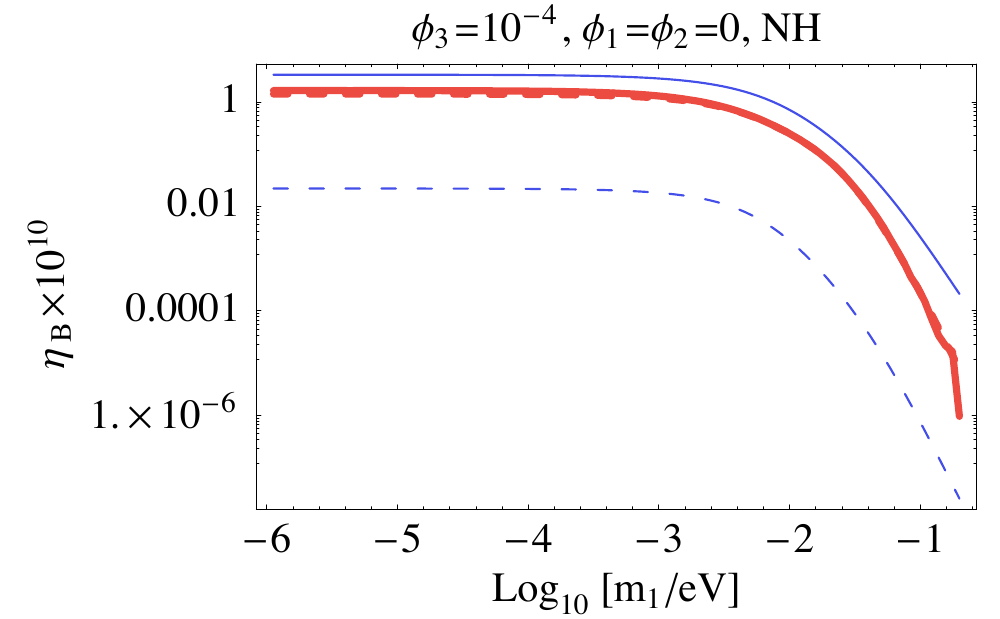}\\
\caption{Dependence on $m_1$, in normal hierarchy. Thick red and thin blue lines  correspond to $c^{(1)}=6\times 10^{-3}$  and $c^{(1)}=2\times 10^{-5}$, respectively.
{\it Top:} The absolute value of the ratio of the  baryon asymmetry with quantum effects 
($\eta_B^{\rm qu}$) and without quantum effects ($\eta_B^{\rm cl}$).
{\it Bottom: } The absolute values of $\eta_B^{\rm qu}$  (solid lines) and $\eta_B^{\rm cl}$ (dashed lines).}\label{fig-NH-m1}
\end{figure}

\begin{figure}[h!]
\vskip 2cm
\includegraphics[scale=0.5]{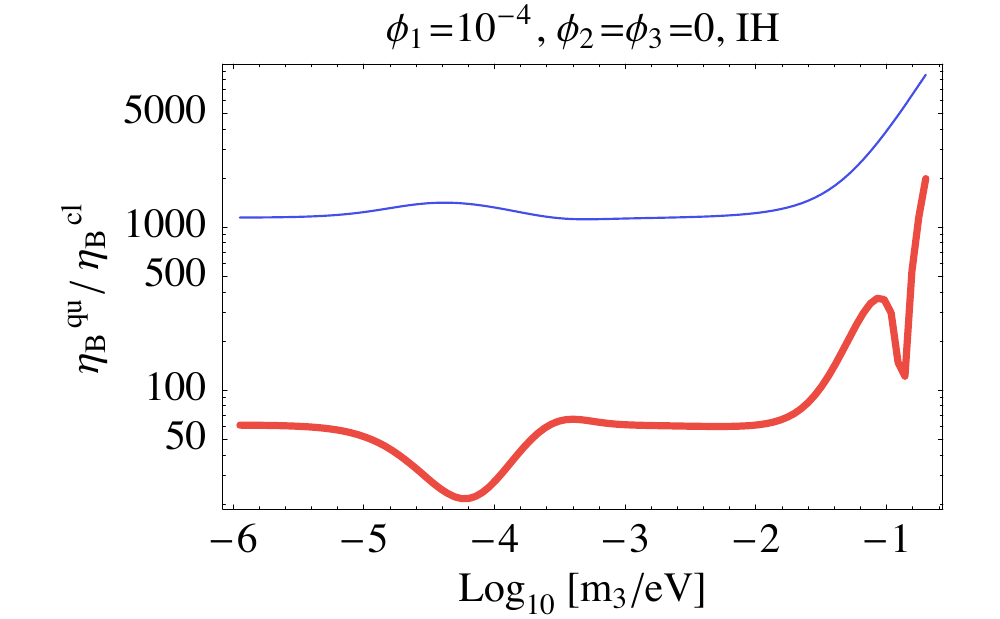}
\includegraphics[scale=0.5]{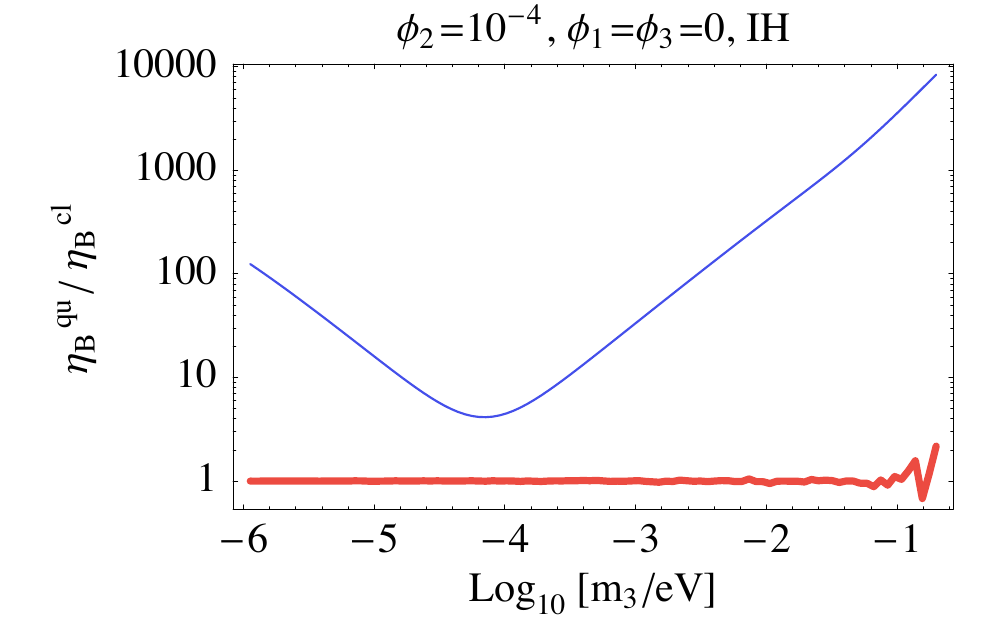}
\includegraphics[scale=0.5]{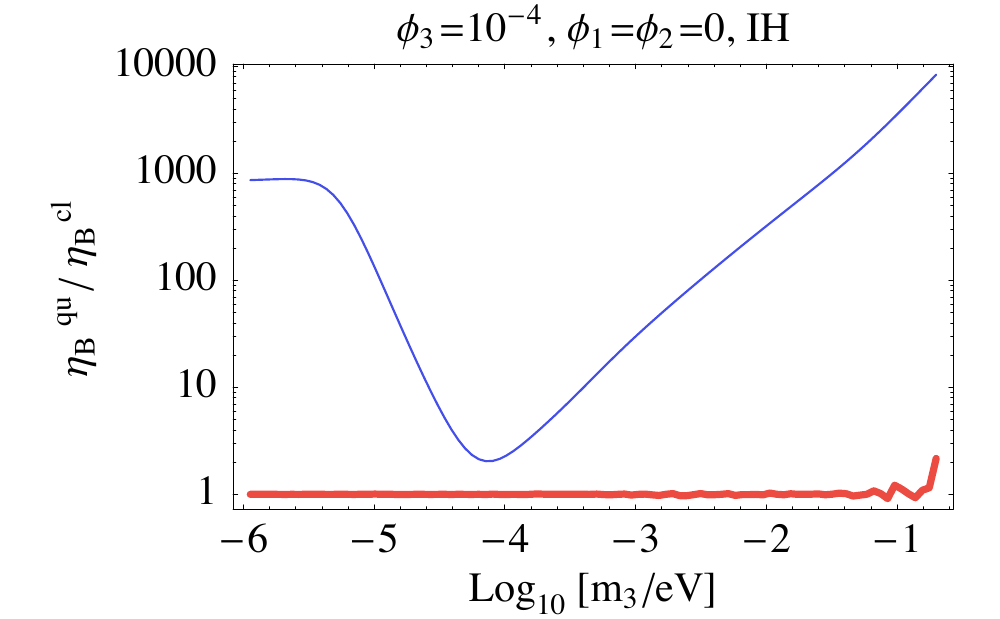}\\
\vskip5pt
\includegraphics[scale=0.5]{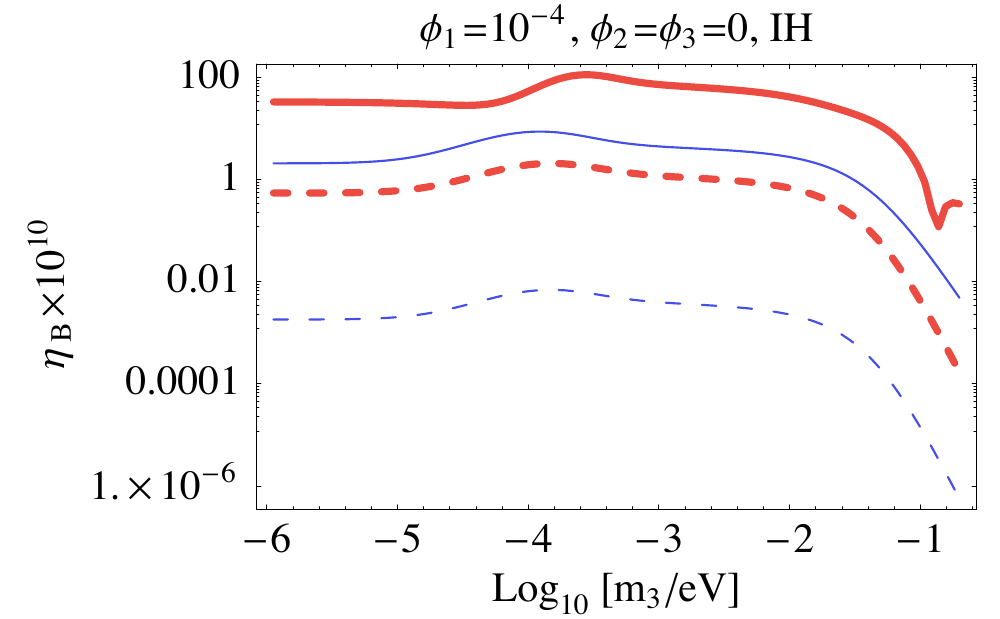}
\includegraphics[scale=0.5]{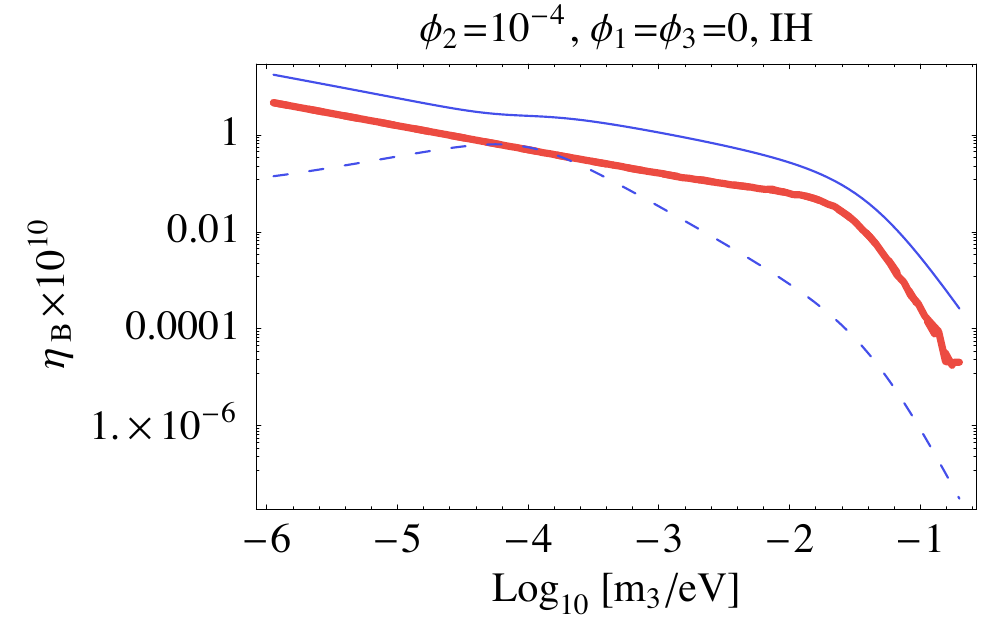}
\includegraphics[scale=0.5]{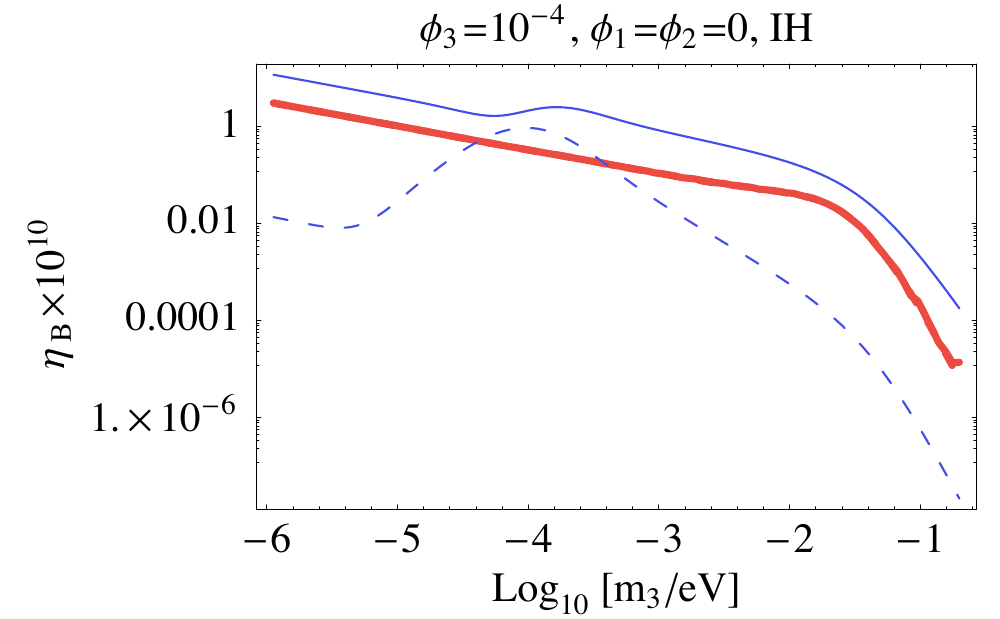}\\
\caption{Dependence on $m_3$, in inverse hierarchy. Thick red and thin blue lines  correspond to $c^{(1)}=6\times 10^{-3}$  and $c^{(1)}=2\times 10^{-5}$, respectively.
{\it Top:} The absolute value of the ratio of the  baryon asymmetry with quantum effects ($\eta_B^{\rm qu}$) and without quantum effects ($\eta_B^{\rm cl}$).
{\it Bottom: } The absolute values of $\eta_B^{\rm qu}$  (solid lines) and $\eta_B^{\rm cl}$ (dashed lines).}\label{fig-IH-m1}
\end{figure}

In order to explore numerically the dependence on $c^{(1)}$, we set $c^{(2)}_{i}=0$ and choose $M_\nu^0=10^9$ GeV. 
Then, we switch on in turn each $\phi_i=10^{-4}$, setting to zero the two others.
Because of the smallness of the $\phi_i$, we will have $\hat m_\nu\approx \hat m_\nu^0$ and $U \approx U^0$; 
for definiteness, we select the parameters of $U^0$ to be: $\theta_{23}=\pi/4$, $\theta_{12}=35^\circ$, 
$\theta_{13}=10^{-3}$, $\delta=\alpha_1=\alpha_2=0$. 

For $m_l=10^{-3}$ eV (thick red line) and $m_l=10^{-1}$ eV (thin blue line),
the upper plots of Figs.~\ref{fig-NH-c1} and \ref{fig-IH-c1} show the ratio of the asymmetry computed 
including quantum effects ($\eta_B^{\rm qu}$) to the classical asymmetry ($\eta_B^{\rm cl}$),
for NH and IH respectively.
The corresponding lower plots show the absolute value of $\eta_B$ with (solid line) and without (dashed line) quantum effects. 
In the classical case, the resonant point is usually reached for $c^{(1)} \sim 0.01$, 
the only exception being IH with $\phi_1\neq 0$ where it is reached for $c^{(1)}=\mathcal{O}(1)$.
Let us recall that the natural range for radiatively-induced $c^{(1)}$ is around $1/(4 \pi)^2=6\times 10^{-3}$.
As can be seen, quantum effects induce an enhancement of $\eta_B$ at small values of $c^{(1)}$. Such an enhancement
is caused mainly by the second ($\sin$) term of the function $m^{(i,j)}$.
For both NH and IH quantum effects seem to be important for $c^{(1)} \lesssim 10^{-3}$, 
the only exception being the case of nearly degenerate light neutrinos with $\phi_1$ as 
the main source of CPV in leptogenesis. In the latter case, values of $c^{(1)}$ up to $10^{-2}$ could do the job.  

The figures also show that with degenerate light neutrinos quantum effects are larger, as expected, 
but $\eta_B$ is more suppressed. Since the baryon asymmetry scales linearly with $\phi_i$ one would
need these phases to be larger than about $10^{-4}$ to reproduce the observed value of the
baryon asymmetry.
The dependence on $m_l$ is shown in more detail by the plots of Figs.~\ref{fig-NH-m1} and \ref{fig-IH-m1} for NH and IH
respectively, selecting two representative values of $c^{(1)}$: $c^{(1)}=6\times 10^{-3}$ (radiatively induced) 
and $c^{(1)}=2\times 10^{-5}$.
Notice that the inclusion of the terms proportional to $c^{(2)}_{1}$,  setting for instance 
$c^{(2)}_{1}=(c^{(1)})^2$ in Eq.(\ref{mnu}),  
does not change the above results.

\subsection{Analytical estimates}

In this subsection we will derive a simple analytical estimate for the ratio
\beq
R\equiv\left.{|\eta_B^{\rm qu}|\over |\eta_B^{\rm cl}|}\right\vert_{c^{(1)}\to 0}
= \left.{|\sum_ \alpha (Y_{\Delta_\alpha})^{\rm qu}|\over |\sum_\alpha (Y_{\Delta_\alpha})^{\rm cl}|}\right\vert_{c^{(1)}\to 0}\,,
\eeq
and check that it is  in agreement with the numerical findings.

The dominant contribution to the asymmetry will come from the least washed-out flavour $\beta$, {\it e.g.} the one with the smallest $K_\beta$. Therefore, as a first approximation we take: 
\beq
R\simeq \left.{|(Y_{\Delta_\beta})^{\rm qu}|\over | (Y_{\Delta_\beta})^{\rm cl}|}\right\vert_{c^{(1)}\to 0}\,.
\eeq
In general, $Y_{\Delta_\beta}$ receives three contributions from the out-of-equilibrium decays of the three $N_i$'s. Each of these terms is proportional to the CP asymmetry $\epsilon_{i\beta}$, which 
contains $(\lambda_\nu\lambda_\nu^\dagger)_{i j} m^{(i,j)}$, with $j\neq i$. The expression (\ref{hzero}) for 
$\lambda_\nu\lambda_\nu^\dagger$ tells us that if {\it e.g.} only $\phi_1\neq 0$, the only non-zero off-diagonal  entries  are  those with $(i,j)=(1,2)$ and $(2,1)$. 

In the limit $c^{(1)}\to 0$, the mass splittings $\Delta M_{ji}$ become small (compared to the scale $M_\nu^0$), and so the arguments of the periodic functions in $m^{(i,j)}$ are also small, in the range of $z$ where the lepton asymmetry is generated. The $\sin^2$ term in $m^{(i,j)}$ is completely negligible; instead, the $\sin$ term has in front an amplitude proportional to $1/\Delta M_{ji}$. Therefore, the dependence on $c^{(1)}$ cancels in this  limit and the resulting quantity is
\beq
\label{approxm}
m^{(i,j)}(z)\simeq -{\Gamma_j\over 2H(M_1)}z^2 \simeq -{1\over 2}K_j z^2\,.
\eeq
This disappearance of $c^{(1)}$  explains the plateaux observed in the upper plots of Figs.~\ref{fig-NH-c1} and \ref{fig-IH-c1} in the limit of very small $c^{(1)}$.

Let us consider the case where only $\phi_i\neq 0$. Looking at the expression (\ref{hzero}), we only have two non-zero asymmetries, say $\epsilon_{j\beta}$ and  $\epsilon_{k\beta}$, but we choose the one containing the least wash-out parameter $K_j$, since it corresponds to the most out-of-equilibrium RH neutrino whose decay produces the largest asymmetry; so we just  take 
$\epsilon_{j\beta}(z)\simeq -\bar \epsilon_{j\beta}K_j z^2/2$, where $\bar \epsilon_{j\beta}$ is the CP asymmetry without the memory factor.

Our choices for the lightest neutrino masses ($m_l=10^{-3}$ eV, $10^{-1}$ eV) make all of the $K_i$'s   greater than or of order 1. In such a case, it is a good approximation to take $\de Y_{N_j}/\de z\simeq \de Y_{N_j}^{\rm eq}/\de z= -z^2 \mathcal{K}_1(z)/4g_*$ (here and in the following $\mathcal{K}_1, \mathcal{K}_2$ are the modified Bessel functions of the first and second kind, respectively). Then, an approximate solution to the Boltzmann equations (Eqs.~(\ref{BENi}) and (\ref{BEDelta}) in Appendix \ref{Boltz}) is found using the steepest descent method:
\begin{eqnarray}
(Y_{\Delta_\beta})^{\rm cl}&\simeq& -{1\over 4g_*}\int_0^\infty \de z\, \bar\epsilon_{j\beta} z^2 \mathcal{K}_1(z)
e^{-{K_\beta |A_{\beta\beta}| \over 4}\int_z^\infty \de z' z'^3 \mathcal{K}_1(z')}\\
&\simeq&-   \bar{\epsilon_{j\beta}\over g_*} {1\over  K_\beta |A_{\beta\beta}| \bar z_1}\,,
\label{no-mem-analytic}
\end{eqnarray}
where $\bar z_1$ satisfies the condition: ${K_\beta |A_{\beta\beta}|\over 4}\bar z_1^3 \mathcal{K}_1(\bar z_1)+{3\over \bar z_1}
-{\mathcal{K}_2({\bar z_1})\over \mathcal{K}_1(\bar z_1)}\simeq 0$. A good analytical approximation for $\bar z_1$, valid for $K_\beta\gtrsim 1$, is $\bar z_1\simeq 3.47\, [\log(K_\beta)]^{0.64}$.

Taking into account the quantum correction factor (\ref{approxm}) gives:
\begin{eqnarray}
(Y_{\Delta_\beta})^{\rm qu}&\simeq& -{1\over 4g_*}\int_0^\infty \de z\, \epsilon_{j\beta}(z) z^2 \mathcal{K}_1(z)
e^{-{K_\beta |A_{\beta\beta}|\over 4}\int_z^\infty \de z' z'^3 \mathcal{K}_1(z')}\\
&\simeq&  {\bar\epsilon_{j\beta}\over 8g_*} K_j\int_0^\infty \de z\,  z^4 \mathcal{K}_1(z)
e^{-{K_\beta |A_{\beta\beta}|\over 4}\int_z^\infty \de z' z'^3 \mathcal{K}_1(z')}\\
&\simeq&   {\bar\epsilon_{j\beta}\over 2g_*} {K_j\over K_{\beta} |A_{\beta\beta}|} \bar z_2\,.
\label{mem-analytic}
\end{eqnarray}
where $\bar z_2$ satisfies the condition: ${K_\beta |A_{\beta\beta}|\over 4}\bar z_2^3 \mathcal{K}_1(\bar z_2)+{4\over \bar z_2}
-{\mathcal{K}_2({\bar z_2})\over \mathcal{K}_1(\bar z_2)}\simeq 0$. Since one expects $\bar z_{1,2}>1$, they satisfy almost the same condition; thus it is reasonable to take: $\bar z_1\simeq \bar z_2\equiv \bar z$.
The expressions (\ref{no-mem-analytic}) and (\ref{mem-analytic}) lead to the simple estimate:
\beq
R(\phi_i\neq 0)\simeq {1\over 2} K_j \bar z^2\sim 10\, K_j \,[\log(K_\beta |A_{\beta\beta}|)]^{1.28}\,,
\label{R-general}
\eeq
where we neglected $\mathcal{O}(1)$ factors, since we are only interested in getting the order of magnitude of the importance of the quantum effects over the classical approximation. 
If we had taken lower values of $m_l$, thus giving  $K_j=\min(K_i)<1$, the analytical estimate would have proceeded in a different way. Though,  $R$ receives a leading contribution  still proportional to $K_j$ as in (\ref{R-general}), plus corrections going like ${1\over z_{\rm eq}^2}{K_\beta\over K_j}$, where $z_{\rm eq}$ is defined as the ``time'' at which the number density of $N_j$ reaches the equilibrium one: $Y_{N_j}(z_{\rm eq})=Y_{N_j}^{\rm eq}(z_{\rm eq})$. For the reference values we considered, this additional term is $\sim 10^{-2} K_\beta/K_j$. As described later, such a correction might be important in the limit of very small $K_j$, {\it i.e.} very small $m_l$, since the leading term is suppressed in that limit.

Let us now apply the analytical result (\ref{R-general}) to the specific cases analysed numerically in the previous subsection.

\paragraph{Normal Hierarchy.}
With the parameters chosen for the numerical analysis, the least washed-out flavour is $\beta=e$, while the $K_i$'s are in the order $K_1<K_2<K_3$. Thus, the previous arguments lead to
\beq
R(\phi_1\neq 0)\simeq R(\phi_2\neq 0)\sim 10\, K_1 \,[\log(K_e |A_{ee}|)]^{1.28}\,,\quad 
R(\phi_3\neq 0)\sim 10\, K_2 \,[\log(K_e |A_{ee}|)]^{1.28}\,.
\eeq
In the case $m_1^0=10^{-3}$ eV, one has $K_1\simeq 0.9, K_2\simeq 8.4$ and $K_e\simeq 3.4$, giving
\beq
R(\phi_1\neq 0)\simeq
R(\phi_2\neq 0)\sim \mathcal{O}(10)  \,,\qquad 
R(\phi_3\neq 0)\sim \mathcal{O}(10^2)  \,.
\eeq
In the case $m_1^0=10^{-1}$ eV, one has $K_1\simeq 93.5, K_2\simeq 93.8$ and $K_e\simeq 93.6$, giving
\beq
R(\phi_1\neq 0)\simeq
R(\phi_2\neq 0)\simeq R(\phi_3\neq 0)\sim \mathcal{O}(10^4)  \,.
\eeq
Such estimates are in good agreement with the numerical values for $\eta_B^{\rm qu}/\eta_B^{\rm cl}$ in the $c^{(1)}\to 0$ limit, shown in the upper plots of Fig.~\ref{fig-NH-c1}.

\paragraph{Inverse Hierarchy.}
In this case the  least washed-out flavour is $\beta=\mu$, while the $K_i$'s are in the order $K_3<K_1<K_2$. Repeating the same analysis as before, one finds
\beq
R(\phi_1\neq 0)\sim 10\, K_1 \,[\log(K_\mu |A_{\mu\mu}|)]^{1.28}\,,\quad 
 R(\phi_2\neq 0)\simeq  R(\phi_3\neq 0)\sim 10\, K_3 \,[\log(K_\mu |A_{\mu\mu}|)]^{1.28}\,.
\eeq
In the case $m_3^0=10^{-3}$ eV, one has $K_1\simeq 46, K_3\simeq 0.9$ and $K_\mu \simeq 24$, giving
\beq
R(\phi_1\neq 0)\sim \mathcal{O}(10^3)     \,,\qquad 
R(\phi_2\neq 0)\simeq 
R(\phi_3\neq 0)\sim \mathcal{O}(10^2)  \,.
\eeq
In the case $m_3^0=10^{-1}$ eV, one has $K_1\simeq 104, K_2\simeq 93$ and $K_\mu\simeq 10^2$, giving
\beq
R(\phi_1\neq 0)\simeq R(\phi_2\neq 0)\simeq R(\phi_3\neq 0)\sim \mathcal{O}(10^4) \,.
\eeq
Again, these orders of magnitudes agree very well with  the $c^{(1)}\to 0$ limit found by numerical integration, as one can see from  the upper plots of  Fig.~\ref{fig-IH-c1}.

As anticipated before, the same analytical approximation (\ref{R-general}) 
qualitatively explains also part of the dependence on 
$m_l$, for very small $c^{(1)}$ (the thin blue line in the upper plots of 
Figs.~\ref{fig-NH-m1} and \ref{fig-IH-m1}).
Indeed, by considering the various $K_\alpha$ and $K_i$ as functions of the light neutrino masses,  as given by Eqs.~(\ref{Ke})-(\ref{Ki}), one can recover the decreasing behaviour of $R$ as $m_l$ becomes small.  In some cases, though, the main term in $R$   given by (\ref{R-general}) turns out  to be proportional to $m_l$ itself. They are the cases $\phi_1,\phi_2\neq 0$ for NH and $\phi_2,\phi_3\neq 0$ for IH. In such situations, when $m_l$ is too small (compared to $m_*\simeq 10^{-3}$ eV) the leading contribution to $R$ is  suppressed and the next-to-leading term proportional to $1/K_j$ starts being important. This describes the change of behaviour which can be seen in the upper plots of Figs.~\ref{fig-NH-m1} and \ref{fig-IH-m1}, in the cases we mentioned above. 

The fact that such a change in $R$ occurs at different values of $m_l$ for NH and IH can be seen as follows. For NH, the next-to-leading order term contains $K_e/K_1$, as functions of the light neutrino 
masses; since $s_{13}$ is tiny, the main contribution will come from the term with $m_2/m_1=m_i/m_l$. On the contrary, for IH the largest term in the ratio $K_\mu/K_3$ is the one with $m_1/m_3=m_h/m_l$, which is generally an order of magnitude greater than $m_i/m_l$ for NH.  Therefore,  the next-to-leading term starts being relevant, in IH, at a larger value of $m_l$ than in NH.

In this Section 
we have considered the effect of the splitting induced by $c^{(1)}$ - namely $\phi_i$ as sources of CPV.
We have found that what is important for quantum effects to be sizable is not the magnitude of the washout parameters $K$, 
which is independent of $c^{(1)}$ in the MLFV framework.
Rather, what is crucial is the frequency and the amplitude of the $\sin$ oscillating term,
which have to be respectively small and large. 
This requires small $c^{(1)}$ and/or strong degeneracy among
the light neutrinos ${\bf m}_j$ and ${\bf m}_i$ for the quantum effects to be sizeable.
In the next Section we will analyse the case in which $H=I$.

%%%%%%%%%%%%%%%%%%%%%%%%%%%%%%%%%%%%%%%%%%%%%%%%%%%%%%%%%%%%%%%%%%%%%%

%\section{Second order splitting from charged leptons ($H=1)$.}
\section{CP violation from light neutrino mixing  ($H=I)$.}
\label{sect:H1}

%%%%%%%%%%%%%%%%%%%%%%%%%%%%%%%%%%%%%%%%%%%%%%%%%%%%%%%%%%%%%%%%%%%%%%

MLFV-leptogenesis  is  viable in the limit $H=I$ only 
in the ``flavoured" regime, 
provided that the heavy neutrino mass matrix receives 
a radiative splitting proportional 
to the charged-lepton Yukawa couplings
($\delta M_\nu= c_{4}^{(2)} \, M_\nu^0  \,  (h_e + h_e^T)$). 
Here we derive the analytic dependence of the asymmetries $\epsilon_{i \alpha}$
on the parameters of the model.

The CP asymmetries read 
 \beq 
 \epsilon_{i \alpha} = \sum_{j\neq i}  \, F^{(j,i)}  \, \frac{  
  {\rm Im} \left(     ( \lambda_\nu  )_{i \alpha}   \,  ( \lambda_\nu^\dagger)_{\alpha j} \,   
 (\lambda_\nu  \lambda_\nu^\dagger)_{ij} \right)	}{| \lambda_\nu \lambda_\nu^\dagger |_{ii} }~,
 \eeq
where, following the notation of Section~3, 
$\lambda_\nu = \bar{U} \lambda_\nu^0$ and 
$\bar{U} M_\nu \bar{U}^T = \hat{M}_\nu$. 
In the limit $\bar{U} \to 1$, where $\lambda \to \lambda_\nu^0$
(see  Eq.~(\ref{la0nu})),
it is easy to verify that the $\epsilon_{i \alpha}$ vanish.  
The key ingredient in obtaining a non-zero result  is to 
have  non-diagonal entries in $h_\nu \equiv \lambda_\nu \lambda_\nu^\dagger $
and correspondingly a non trivial $\bar{U}$ matrix.
In order to estimate the off-diagonal entries of $h_\nu$ and $\bar{U}$ we 
perform a perturbative expansion of $M_\nu$
assuming $c^{(2)}_{i} \ll c^{(1)}$: 
\bea
M_\nu  &=& M_\nu^0  
\Bigg\{I + 2 c^{(1)} \, h_\nu^0  +  (2  c^{(2)}_{1} + c_{2}^{(2)} + c_{3}^{(2)} ) \, (h_\nu^0)^2   + c_{4}^{(2)} \,
(h_e + h_e^T) + \cdots \Bigg\}~,  
\\
&=&  M + \Delta~.
\eea
Here $M$ is a real diagonal matrix, 
with eigenvalues $M_i = M_\nu^0  [1 + 2 c^{(1)} {\bf m}_i/\tilde{m} + 
(2 c_{1}^{(2)} + c_{2}^{(2)} + c_{3}^{(2)} ) ({\bf m}_i/\tilde{m})^2]$, 
and $\Delta$ is a real symmetric matrix given by 
\beq
\Delta =  
 \left\{
\begin{array}{ll} 
 c_{4}^{(2)}  M_\nu^0  \, (h_e + h_e^T)   &   {\rm  NH}~, \\ 
 &  \\
c_{4}^{(2)}  M_\nu^0  \, \tilde{I} \, (h_e + h_e^T) \, \tilde{I}^T    &   {\rm  IH}~,
\end{array} \right. 
\eeq
with the $\tilde{I}$ defined in Eq.~(\ref{zz}). 
$M_\nu$ is diagonalised perturbatively by the real orthogonal matrix $\bar{U} = I + T$
($T^T = - T$), with
\beq
T_{ij} =  \frac{\Delta_{ij}}{M_i - M_j}~. 
\eeq 
To first order in $c_{4}^{(2)}/c^{(1)}$ the explicit 
epression for 
$\lambda_\nu  \lambda_\nu^\dagger = \bar{U} h_\nu^0 \bar{U}^\dagger$
is then
\beq 
(\lambda_\nu \lambda_\nu^\dagger)_{i \neq j} = \frac{\Delta_{ij} \, ({\bf m}_j - {\bf m}_i)}{\tilde{m} (M_i - M_j)} 
 =  -  \frac{ c_{4}^{(2)}}{c^{(1)}}   \times \left\{
\begin{array}{ll} 
  {\rm Re}  (h_e)_{ij}   &   {\rm  NH}~, \\ 
 &  \\
 \, {\rm Re}  (\tilde{I}  \, h_e \, \tilde{I}^T)_{ij}   &   {\rm  IH}~.
\end{array} \right. 
\eeq 
Using the above result in the expression of $\epsilon_{i \alpha}$ (along with 
$(\lambda_\nu^0)_{i \alpha} (\lambda_\nu^0)^\dagger_{\alpha j}$), one obtains for NH:
\beq
\epsilon_{i \alpha}^{NH} =  -\frac{c_{4}^{(2)} }{c^{(1)}} \, \sum_{\beta=e,\mu,\tau} \left( \frac{m_\beta}{v}
\right)^2 \, 
\sum_{j \neq i}   \Bigg\{ 
F^{(j,i)} \ \frac{{ m}_j}{\tilde{m}}  \  {\rm Re} \left(  U_{\beta i}^* U_{\beta j} \right)  
\  {\rm Im} \left(  U_{\alpha i}^* U_{\alpha j} \right)  
\Bigg\} ~,
\eeq
while for IH:
\beq
\epsilon_{i \alpha}^{IH} =  - \frac{c_{4}^{(2)} }{c^{(1)}} \, \sum_{\beta=e,\mu,\tau}
 \left( \frac{m_\beta}{v} \right)^2 \, 
\sum_{j \neq i}  \sum_{m,n}  \Bigg\{ 
F^{(j,i)} \  \frac{{ m}_n}{\tilde{m}} \  \tilde{I}_{im} \tilde{I}_{jn}
{\rm Re} \left(  U_{\beta m}^* U_{\beta n} \right)  
\  {\rm Im} \left(  U_{\alpha m}^* U_{\alpha n} \right)  
\Bigg\} ~.
\eeq
In the above expressions the dominant term arises when 
$\beta=\tau$ in the sum over charged lepton flavours. Keeping only 
$\beta=\tau$ and defining 
\beq
\Phi_{ij}^{(\alpha)} \equiv   {\rm Re} \left(  U_{\tau i}^* U_{\tau j} \right)  
\  {\rm Im} \left(  U_{\alpha i}^* U_{\alpha j} \right) ~, 
\eeq
the asymmetries read:
\bea
\epsilon_{1 \alpha}^{NH} &=&  -  \frac{c_{4}^{(2)} }{c^{(1)}} \,  \left( \frac{m_\tau}{v} \right)^2 \, 
\left[ 
 F^{(2,1)}  \frac{{ m}_2}{\tilde m}  \, \Phi_{12}^{(\alpha)}   
+   F^{(3,1)}  \frac{{ m}_3}{\tilde m}  \, \Phi_{13}^{(\alpha)}    
\right]
\\
\epsilon_{2 \alpha}^{NH} &=&   + \frac{c_{4}^{(2)} }{c^{(1)}} \,  \left( \frac{m_\tau}{v} \right)^2 \, 
\left[ 
 F^{(1,2)}  \frac{{ m}_1}{\tilde m}  \, \Phi_{12}^{(\alpha)}   
-  F^{(3,2)}  \frac{{ m}_3}{\tilde m}  \, \Phi_{23}^{(\alpha)}    
\right]
\\
\epsilon_{3 \alpha}^{NH} &=&   + \frac{c_{4}^{(2)} }{c^{(1)}} \,  \left( \frac{m_\tau}{v} \right)^2 \, 
\left[ 
 F^{(1,3)}  \frac{{ m}_1}{\tilde m}  \, \Phi_{13}^{(\alpha)}   
+   F^{(2,3)}  \frac{{ m}_2}{\tilde m}  \, \Phi_{23}^{(\alpha)}    
\right] ~. 
\eea
The IH case is obtained by a straightforward permutation of indices in the 
${ m}_j$ and $\Phi_{ij}^{\alpha}$ factors.
We are now in a position to identify the dependence of the baryon asymmetry on the 
low-energy CP violating phases, contained in  the factors $\Phi_{ij}^{(\alpha)}$.  
In the case of NH, only  $\Phi_{ij}^{(e)}$   are relevant, as the $e$ flavour is the least washed-out one:
\bea
\Phi_{12}^{(e)} &=&  \frac{1}{2} \  s_{12}^2 c_{12}^2  s_{23}^2  c_{13}^2 \,  \sin (\alpha_1- \alpha_2) 
+\cO(s_{13}^2),
\\
\Phi_{13}^{(e)} &=& \  s_{12} c_{12}  s_{23} c_{23}  c_{13}^2  \,  s_{13}  
\,   \cos \left(\frac{\alpha_1}{2} \right)  \,   \sin \left(\frac{\alpha_1}{2} + \delta \right)  
+\cO(s_{13}^2),
\\
\Phi_{23}^{(e)} &=& \  s_{12} c_{12}  s_{23} c_{23}  c_{13}^2 \,  s_{13}  
\,    \cos \left(\frac{\alpha_2}{2} \right)  \   \sin \left(\frac{\alpha_2}{2} + \delta \right)
+\cO(s_{13}^2)~.
\eea
For IH, all flavours are roughly equally  washed out, so one needs also 
$\Phi_{ij}^{(\mu,\tau)}$. Expanding to first non-trivial order in $s_{13}$ 
we find:
\bea
\Phi_{12}^{(\mu)} &=& -  \frac{1}{2} \  s_{12}^2 c_{12}^2  s_{23}^2  c_{23}^2 \,  \sin (\alpha_1- \alpha_2)~, 
\\
\Phi_{13}^{(\mu)} &=& \frac{1}{2}  \  s_{12}^2 s_{23}^2  c_{23}^2   c_{13}^2  
\, \sin \alpha_1~,  \\
\Phi_{23}^{(\mu)} &=& \frac{1}{2}  \   c_{12}^2 s_{23}^2  c_{23}^2   c_{13}^2 
\,    \sin \alpha_2  ~,
\\
%\eea
%\bea
\Phi_{12}^{(\tau)} &=& -  \frac{1}{2} \  s_{12}^2 c_{12}^2  s_{23}^4  \,  \sin (\alpha_1- \alpha_2)~,
\\
\Phi_{13}^{(\tau)} &=& - \frac{1}{2}  \  s_{12}^2 s_{23}^2  c_{23}^2   c_{13}^2   
\, \sin \alpha_1~,  \\
\Phi_{23}^{(\tau)} &=& - \frac{1}{2}  \   c_{12}^2 s_{23}^2  c_{23}^2   c_{13}^2 
\,    \sin \alpha_2  ~. 
\eea

The above expressions explain well the two main qualitative features 
of this scenario:
\begin{itemize}
\item{} In order to have a lepton asymmetry which does not vanish in the limit 
$s_{13} \to 0$ it is necessary to have non-vanishing Majorana phases.
\item{} In the limit $\alpha_1 =\alpha_2 =0$ the lepton asymmetry
is proportional to $ s_{13} \sin \delta$
but the overall scale is substantially smaller than in the 
generic case with non-vanishing Majorana phases. 
\end{itemize}

\begin{figure}[t!]
\begin{center}
\includegraphics[scale=0.9]{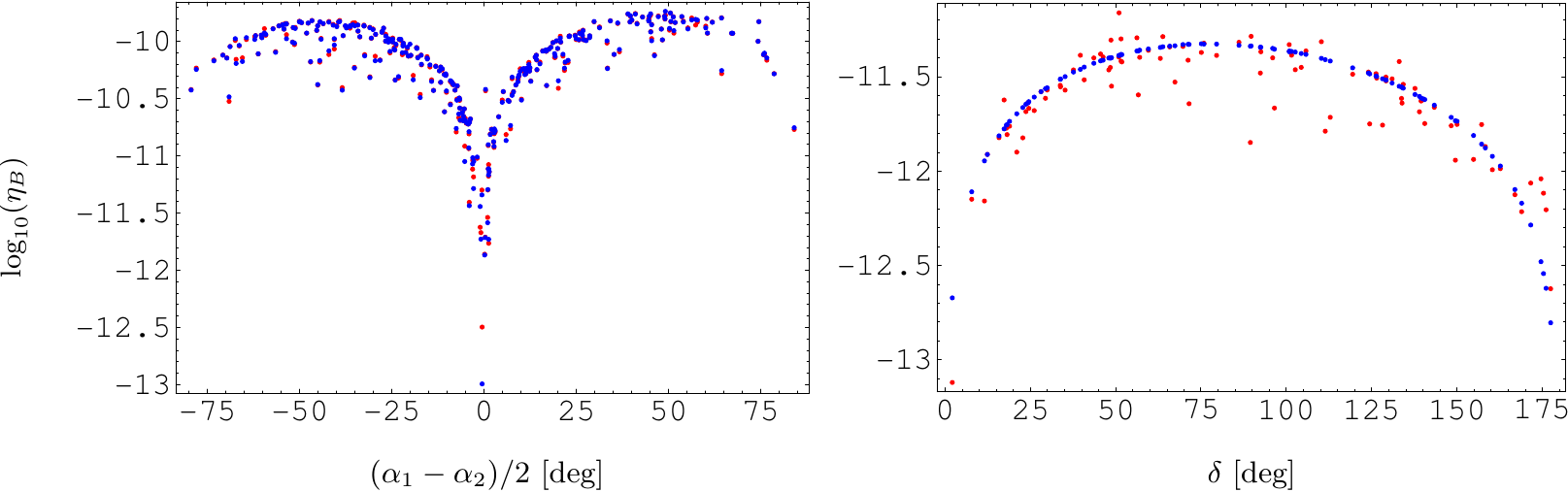} 
\vskip -0.5 cm 
\end{center}
\caption{{\it Left:} $\eta_B$ vs. $(\alpha_1-\alpha_2)/2$  varying 
$m_1\leq 10^{-2}$~eV, $s_{13} \leq 0.1$, and $0< \delta < 180^\circ$
in the NH scenario. {\it Right:}  $\eta_B$ vs. $\delta$  for
$\alpha_1=\alpha_2=0$, setting 
$m_1 = 10^{-3}$~eV and $s_{13}= 0.03$  in the NH scenario.
In both plots $c_{4}^{(2)} =10^{-2}$, $c^{(1)}=10^{-1}$, $M_\nu=10^9$~GeV,
and red (blue) points correspond to the inclusion (exclusion) 
of quantum effects.
\label{fig:H1} }
\vskip 0.7 cm 
\end{figure}
\begin{figure}[t!]
\begin{center}
\includegraphics[scale=0.6]{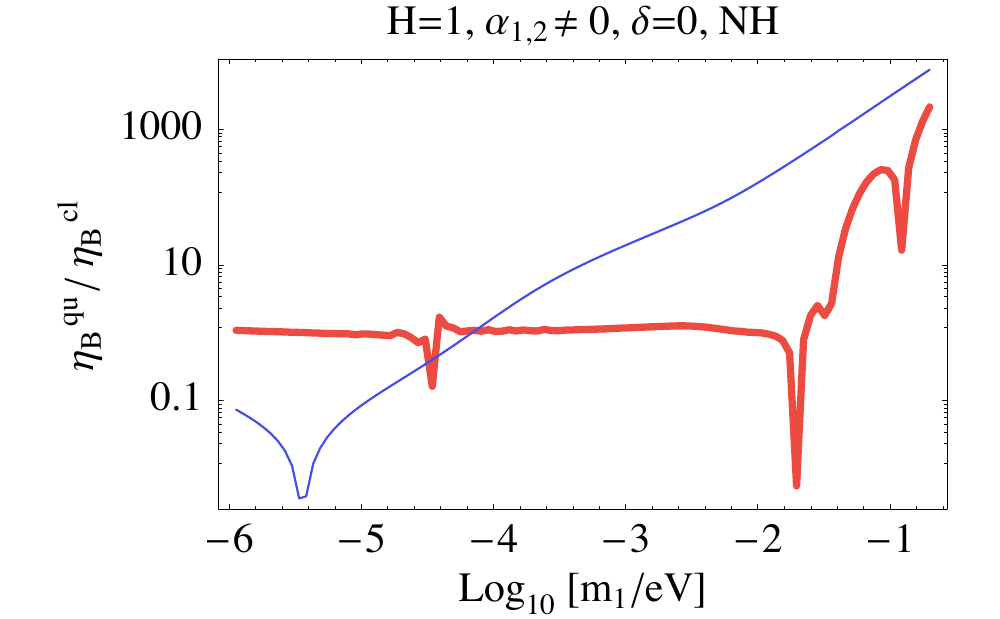}
\includegraphics[scale=0.6]{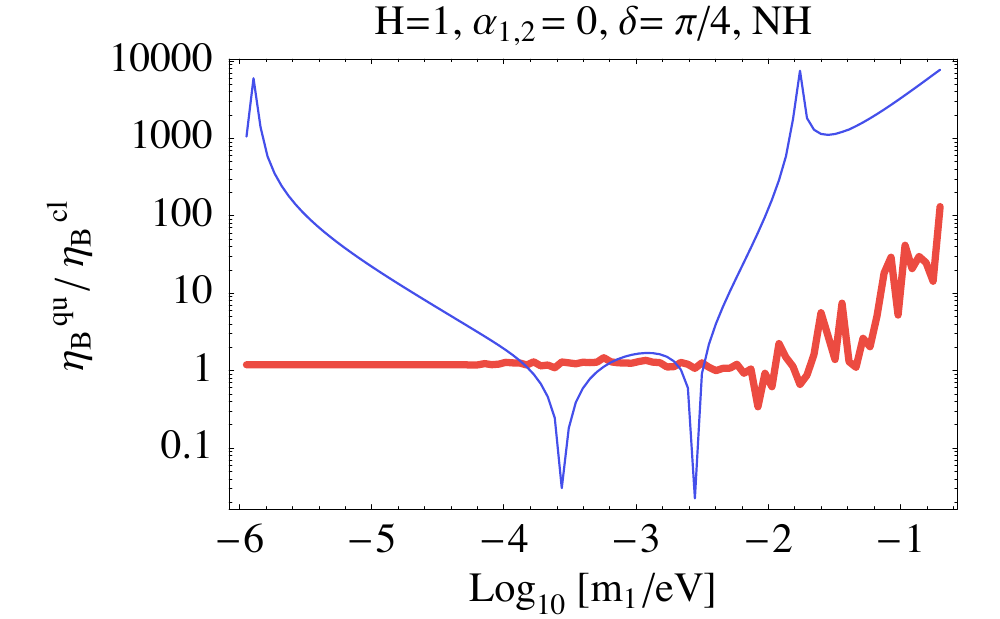}\\
\vskip5pt
\includegraphics[scale=0.6]{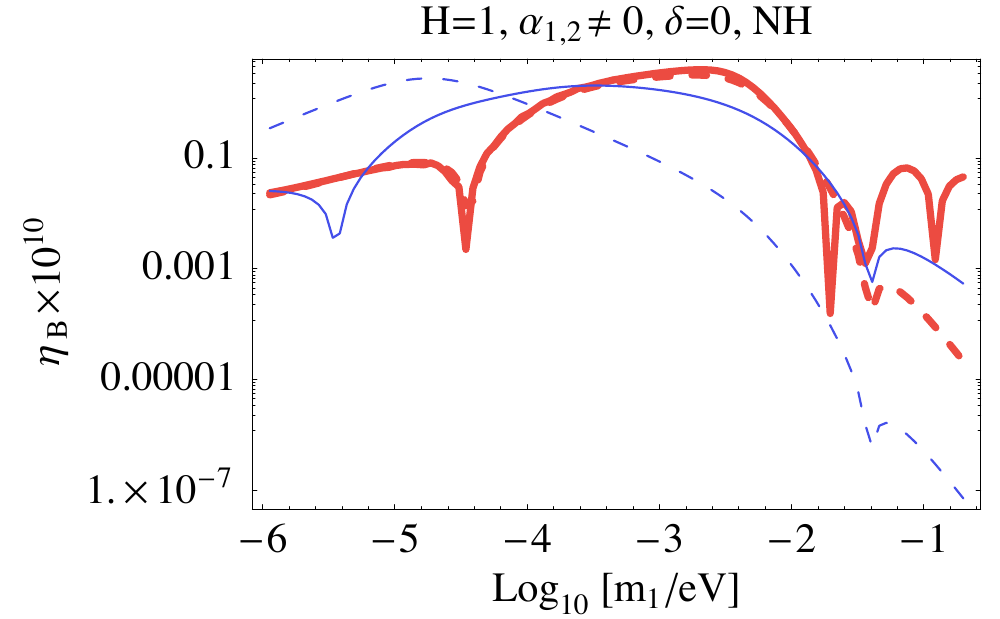}
\includegraphics[scale=0.6]{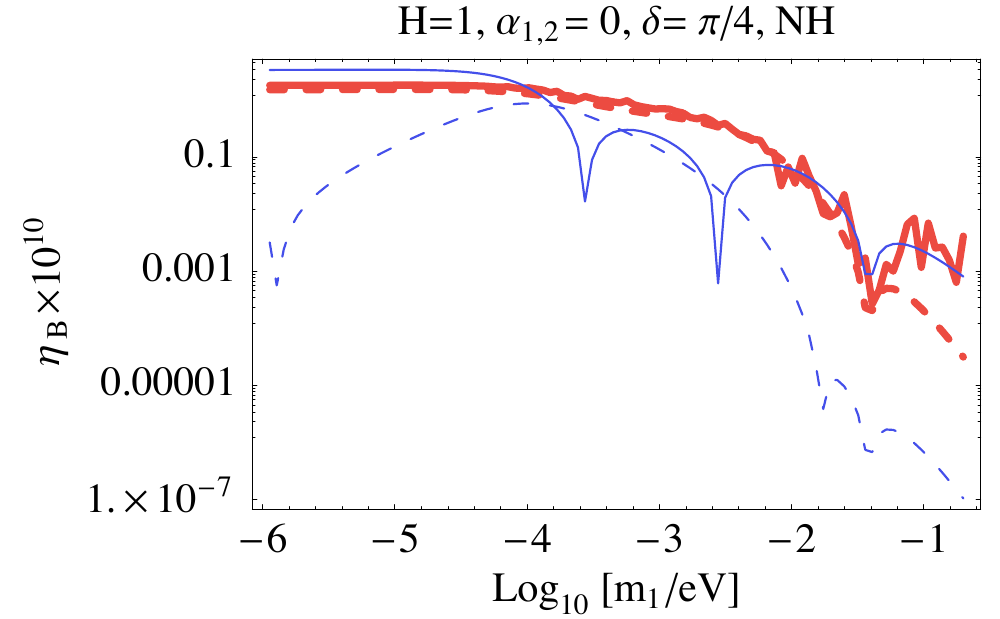}\\
\end{center}
\caption{Dependence on $m_1$, in normal hierarchy and $H=1$. Thick red and thin blue lines  correspond to $c^{(1)}=6\times 10^{-3}$  and $c^{(1)}=2\times 10^{-5}$, respectively; the other RH mass splitting coefficients are taken to be: $c^{(2)}_{1,2,3}=0$ and  $c^{(2)}_4=0.1 c^{(1)}$. The PMNS phases and $s_{13}$ are chosen to be $\alpha_1=36^\circ, \alpha_2=60^\circ, \delta =0, s_{13}=10^{-3}$  and 
$\alpha_1= \alpha_2=0, \delta =\pi/4, s_{13}=0.1$.
{\it Top:} The absolute value of the ratio of the  baryon asymmetry with quantum effects ($\eta_B^{\rm qu}$) and without quantum effects ($\eta_B^{\rm cl}$).
{\it Bottom: } The absolute values of $\eta_B^{\rm qu}$  (solid lines) and $\eta_B^{\rm cl}$ (dashed lines). \label{fig:H1m1}}
\end{figure}

\begin{figure}[h!]
\begin{center}
\includegraphics[scale=0.6]{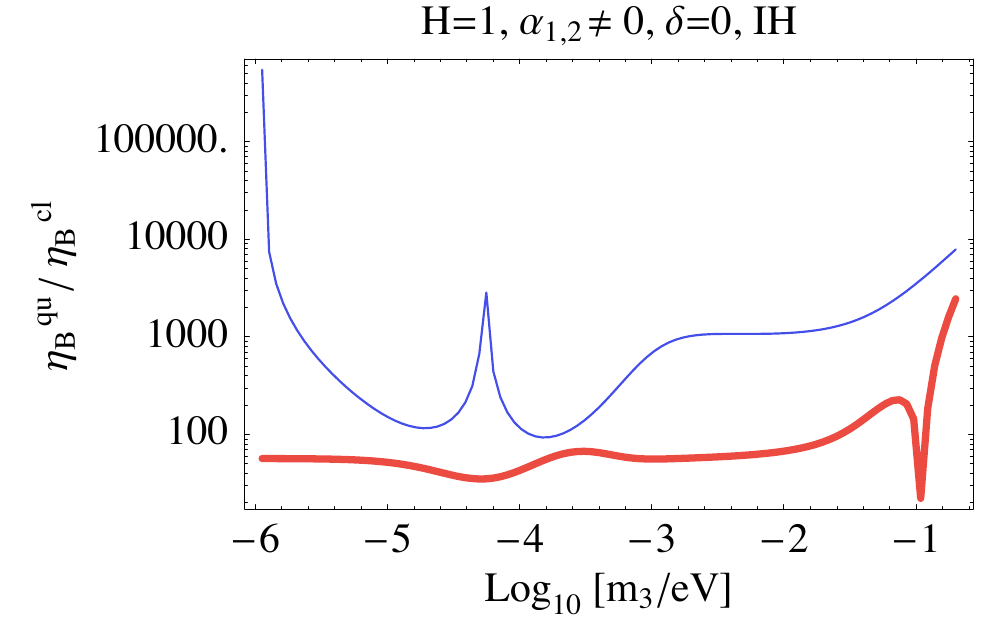}
\includegraphics[scale=0.6]{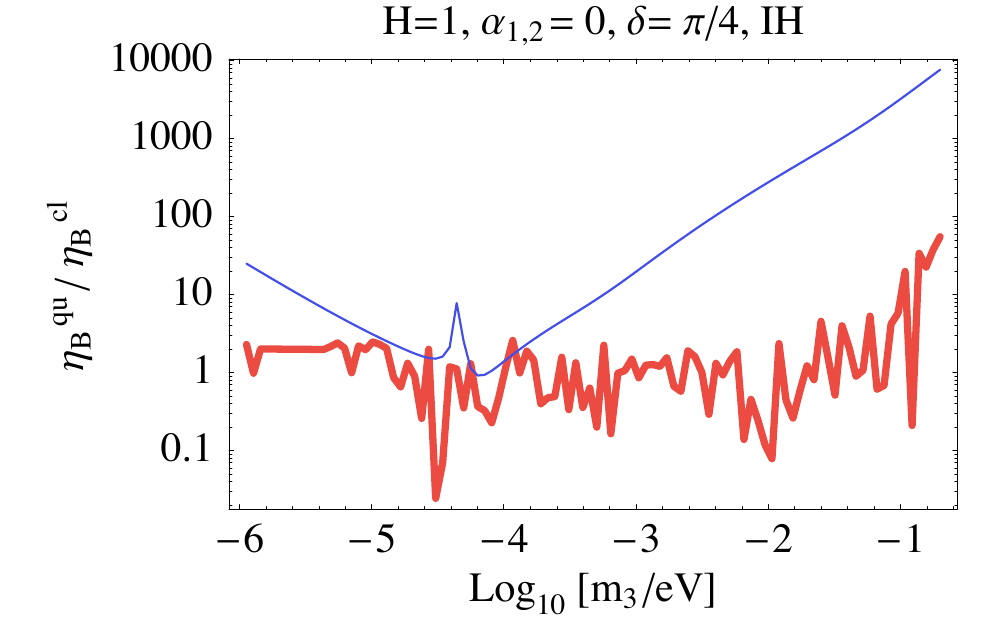}\\
\vskip5pt
\includegraphics[scale=0.6]{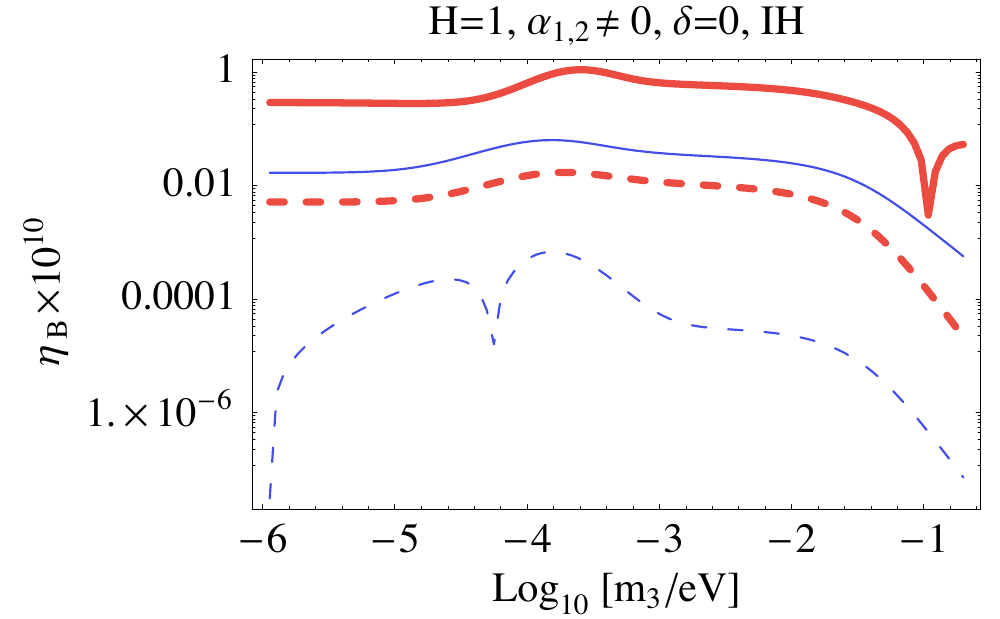}
\includegraphics[scale=0.6]{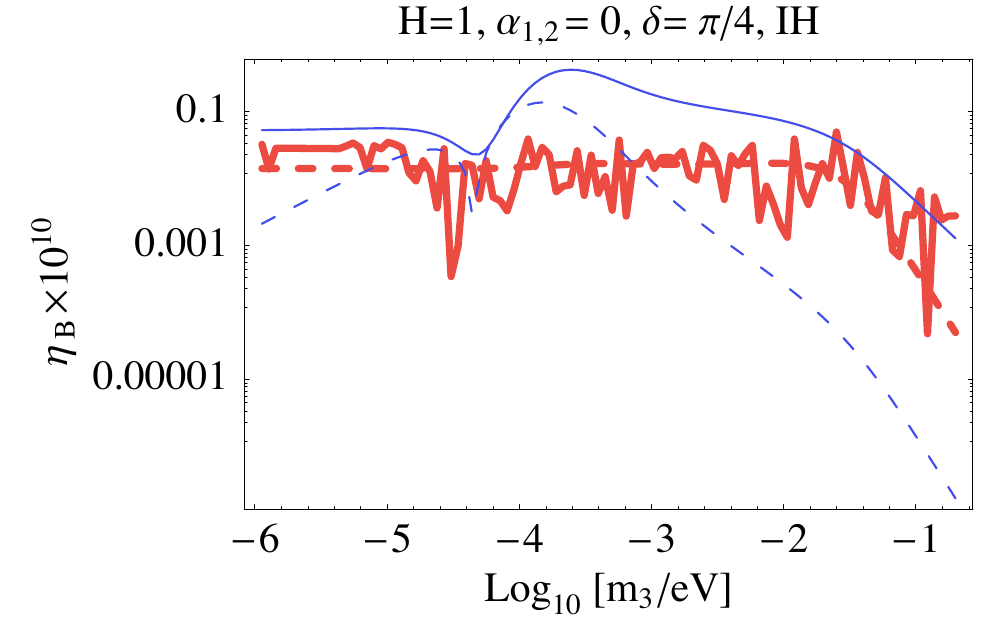}\\
\end{center}
\caption{Dependence on $m_3$, in inverse hierarchy. Thick red and thin blue lines  correspond to $c^{(1)}=6\times 10^{-3}$  and $c^{(1)}=2\times 10^{-5}$, respectively;
 the other RH mass splitting coefficients are taken to be: $c^{(2)}_{1,2,3}=0$ and  $c^{(2)}_4=0.1 c^{(1)}$. The PMNS phases and $s_{13}$ are chosen to be $\alpha_1=36^\circ, \alpha_2=60^\circ, \delta =0, s_{13}=10^{-3}$  and 
$\alpha_1= \alpha_2=0, \delta =\pi/4, s_{13}=0.1$.
{\it Top:} The absolute value of the ratio of the  baryon asymmetry with quantum effects ($\eta_B^{\rm qu}$) and without quantum effects ($\eta_B^{\rm cl}$).
{\it Bottom: } The absolute values of $\eta_B^{\rm qu}$  (solid lines) and $\eta_B^{\rm cl}$ (dashed lines). \label{fig:H1m2}}
\end{figure}

Qualitatively, these two observations agree with the conclusions obtained in 
Ref.~\cite{Uhlig:2006xf}  by means of a numerical study.\footnote{~A precise numerical 
comparison with Ref.~\cite{Uhlig:2006xf} is not possible given the different
assumptions about the structure of the $c^{(i)}_{j}$.}
They are also qualitatively confirmed by our numerical evaluation of the baryon
asymmetry taking into account quantum  effects, as shown in Fig.~\ref{fig:H1}.

It should be stressed that the above observations are only the main 
qualitative features 
and do not describe in detail all the allowed possibilities. In particular, 
the complete numerical study shows that the $\eta_B \propto s_{13} \sin \delta$ 
behaviour in the $\alpha_1 =\alpha_2 =0$ limit, derived from 
a linear expansion in $s_{13}$ and in the approximation 
$ K_e \ll K_{\mu,\tau}$, is a good approximation only for 
$s_{13}\ll 0.1$ and small $m_1$.  For large $s_{13}$ and large $m_1$ values the $\cO(s_{13}^2)$
terms generate a sizable correction to the formally leading 
 $s_{13} \sin \delta$ term. As a result, the asymmetry assumes 
a form of the type $\eta_B \propto  \sin\delta (1 + b \cos\delta)$, 
with $b=\cO(1)$.

As far as the size of the baryon asymmetry is concerned, if the Majorana phases are
large one can easily reach values of $\eta_B$ which are 
consistent with the experimental observations (see Fig.~\ref{fig:H1} left). 
This is not possible  
if the Majorana phases are set to zero, at least in the single-Higgs scenario considered 
so far. In this case $\eta_B$ turns out  to be smaller than the experimental value 
by at least one order of magnitude. However, the overall normalization 
of  $\eta_B$ changes if we consider a two-Higgs doublet scenario (such as the Higgs sector 
of the MSSM), where the charged-lepton Yukawa couplings are enhanced 
by $\tan\beta = \langle H_U\rangle/ \langle H_D \rangle > 1$. In this case 
$\eta_B \propto \tan^2\beta$  and the overall normalization can easily reach 
the experimental value even if the Majorana phases are set to zero. 
The MLFV framework with $H=I$, vanishing Majorana phases and 
large $\tan\beta$,  provides a concrete scenario  where the 
observed baryon asymmetry is directly linked to the measurable 
CP violating phase of the low-energy neutrino mass matrix.

We conclude that the impact of quantum effects in the $H= I$ case  is qualitatively 
very similar to the $H\not = I$ scenario: memory effects become sizable
for small splitting (in particular small $c^{(1)}$) and  
in the limit of light degenerate neutrinos. 
An illustration of the impact of these  effects 
is provided by the the plots in Fig.~\ref{fig:H1m1}--\ref{fig:H1m2}.
As can be noted, the effects can be quite dramatic, especially for IH. 
However, it should also be stressed that the largest relative 
impact is obtained in regions of the parameter space where 
$\eta_B$ is well below the experimental value.

\section{Conclusions}
\label{sect:conclusions}
In this paper we have studied resonant leptogenesis in the MLFV framework where 
it is assumed that 
the 
charged-lepton and the neutrino Yukawa couplings are the 
only irreducible sources of lepton-flavour symmetry 
breaking. In such a framework, the heavy RH neutrinos are highly
degenerate in mass and their decays in the early Universe may give rise to the
observed baryon asymmetry through the mechanism of 
resonant 
leptogenesis. Previous studies on the 
viability of leptogenesis in the MLFV framework ~\cite{Cirigliano:2006nu, Branco:2006hz, Uhlig:2006xf} 
have been based on the assumption that classical Boltzmann equations suffice to analyse the
dynamical generation of the baryon asymmetry. However, it has been recently shown \cite{dsr1,dsr2}
that quantum Boltzmann equations are a more appropriate tool to study such a dynamics when 
the heavy RH neutrinos are degenerate in mass. Indeed, the quantum  Boltzmann equations obtained  
starting from  the  non-equilibrium quantum field theory reveal that the CP asymmetry
is a time-dependent oscillatory function which reduces to the value obtained in the classical approach
only if the oscillation time is much larger than the interaction time. In resonant leptogensis this
is not the case.

We have shown both analytically and numerically that neglecting the time dependence of the
CP asymmetry may underestimate the baryon asymmetry by several orders of magnitude when 
a  strong degeneracy among heavy RH neutrinos and small mass splittings in the light
neutrino sectors are present. This is true both when  the CP phases 
come from the RH sector (phases in the matrix $H$) and when they come entirely from the left-handed
sector  ($H=I$) and may be identified with the low energy PMNS phases.

%%%%%%%%%%%%%%%%%%%%%%%%%%%%%%%%%%%%%%%%%%%%%

\section*{Acknowledgements}
The work of ADS is supported in part by the INFN ``Bruno Rossi'' Fellowship and in part by the US Department of Energy (DOE) under cooperative research agreement DE-FG02-
05ER41360.
The work of GI is supported in part 
by the EU Contract No.~MRTN-CT-2006-035482 {\em FLAVIAnet}.
The work of AR is supported in part by the European Programmes ``The
Quest For Unification'', contract MRTN-CT-2004-503369 and 
 ``UniverseNet", contract MRTN-CT-2006-035863.

%%%%%%%%%%%%%%%%%%%%%%%%%%

\appendix

\section{Conventions for Boltzmann equations}
\label{Boltz}
\noindent 
In the early Universe the quantum numbers conserved by sphaleron interactions are the $\Delta_\alpha=B/3-L_\alpha$. 
The pair of Boltzmann equations describing the generation of the baryon asymmetry are  
\bea
\frac{d Y_{N_i}}{d z} &=& - {D_i}   \left(Y_{N_i}-Y_{N_i}^{eq} \right),
\label{BENi} \\ \nonumber \\
\frac{d Y_{\Delta_\alpha}}{d z} &=& - \sum_i \epsilon_{i\alpha} D_i  \left(Y_{N_i}-Y_{N_i}^{eq} \right) -  
W_\alpha |A_{\alpha\alpha}| Y_{\Delta_\alpha}, 
\label{BEDelta}
\eea
where  at equilibrium the $N_i$ number density normalised to the entropy density of the universe is 
$Y^{eq}_{N_i}= z_i^2 \mathcal{K}_2(z_i)/(4 g_*)$, where $z_i=z \sqrt{x_i}$, $g_*=106.75$ and $\mathcal{K}_2(z_i)$
is a modified Bessel function of the second kind. The washout parameters are generically defined by
\beq
K_{i\alpha}\equiv{\Gamma (N_i \to \ell_\alpha \bar{H})\over H(T=M_i)}\,,
\label{Kialpha}
\eeq
and we also make use of the quantities: $K_\alpha=\sum_i K_{i\alpha}$ and $K_i=\sum_\alpha K_{i\alpha}$. In terms of the parameters of the model, 
$K_{i\alpha} = |(\lambda_\nu)_{i\alpha}|^2 v^2/(M_i m_*)$ with $m_*\approx 10^{-3}$ eV. 
The decay and washout terms appearing in the Boltzmann equations are 
\beq
D_i=K_i~ x_i~ z ~\frac{\mathcal{K}_1(z_i)}{\mathcal{K}_2(z_i)}~~,~~~~ 
W_\alpha= \sum_i \frac{1}{4} K_{i\alpha} ~\sqrt{x_i}~ \mathcal{K}_1(z_i)~ z_i^3~~~,
\eeq
while the matrix A is given by\footnote{We used the 
approximation in which $A$ is a diagonal matrix. For the full expression
see refs.~\cite{f1,f2,f3}.} 
\beq
A=-{\rm diag}(151/179,344/537,344/537),~~~~~ 
{\rm for}~~ M_1 \lesssim 10^{9} {\rm GeV}.
\eeq
For $10^9 {\rm GeV} \lesssim M_1 \lesssim 10^{12} {\rm GeV}$ and $M_1 \gtrsim 10^{12}$ GeV, 
the two-flavour  and the one-flavour regimes should be applied, respectively \cite{f3}.

%Approximate expressions for the Bessel functions are used:
%\beq
%\mathcal{K}_1(z_i)= \frac{1}{z_i} ~\sqrt{1+\frac{\pi}{2}z_i}~ e^{-z_i}~~~,~~~~~
%\mathcal{K}_2(z_i)= \frac{1}{z_i} ~\left(\frac{15}{8}+z_i\right) ~\mathcal{K}_1(z_i)~~. 
%\eeq
%Initial conditions are $N_{N_i}(z_{in})=0$, $N^\alpha_{B-L}(z_{in})=0$ with $z_{in}=.1$.
Finally,
\beq
Y_B=\frac{12}{37}  \sum_\alpha Y_{\Delta_\alpha}(z\rightarrow \infty)~~,
\eeq 
to be compared with the measured value $Y_B = (8.7\pm 0.3) \times 10^{-11}$ or with 
the baryon asymmetry normalized with respect to the photon number density 
(instead of the entropy density) $\eta_B=(6.3\pm 0.3)\times 10^{-10}$.

%%%%%%%%%%%%%%%%%%%%%%%%%%%%%%%%%%%%%%%%%%%%%%%%%%

\section{Analysis of CP violating weak-basis invariants} 
\label{appB}

The independent CP-violating phases of the model can be characterised in terms
of weak-basis invariants, {\it i.e.}~quantities that are insensitive to changes of
basis or re-phasing of the lepton fields.
The  MLFV scenario under investigation has  six  independent
CPV invariants coming from the Yukawa sector.
%Even in the more restrictive case of
%$\lambda_e = 0$ or $\lambda_e = I$  the model contains three independent CPV invariants.

The simplest necessary conditions for CP invariance can
be cast in the following  weak-basis invariant
form~\cite{Branco:2001pq}
\begin{eqnarray}
B_1 & \equiv & {\rm Im} \,
{\rm Tr} \ \left[   h_\nu  \, (M_\nu^\dagger M_\nu)   M_\nu^*   \, h_\nu^*  M_\nu  \right] = 0~,  \\
B_2 & \equiv & {\rm Im} \,
{\rm Tr} \ \left[   h_\nu  \, (M_\nu^\dagger M_\nu)^2   M_\nu^*   \, h_\nu^*  M_\nu  \right] = 0~,  \\
B_3 & \equiv & {\rm Im} \, 
{\rm Tr} \ \left[   h_\nu  \, (M_\nu^\dagger M_\nu)^2   M_\nu^*   \, h_\nu^*  M_\nu (M_\nu^\dagger M_\nu)\right] = 0  \ , 
\label{eq:invariants}
\end{eqnarray}
where
$h_\nu = \lambda_\nu \lambda_\nu^\dagger$
and $M_\nu$ denotes a generic heavy neutrino mass term.
The invariants $B_{1,2,3}$ are independent and survive in the limit $\lambda_e \to 0$. 
One can construct  three other independent invariants that {\it explicitly} \footnote{
Of course $\lambda_e$ can also appear implicitly through the contributions to 
$M_\nu$  allowed by MLFV. }
involve $\lambda_e$ by simply replacing one $h_\nu$ entry in $B_{1,2,3}$  with 
$h_e \equiv \lambda_\nu \lambda_e^\dagger \lambda_e \lambda_\nu^\dagger$:  
\begin{eqnarray}
\tilde{B}_1 & \equiv & {\rm Im} \,
{\rm Tr} \ \left[   h_\nu  \, (M_\nu^\dagger M_\nu)   M_\nu^*   \, h_e^*  M_\nu  \right] = 0~,  \\
\tilde{B}_2 & \equiv & {\rm Im} \,
{\rm Tr} \ \left[   h_\nu  \, (M_\nu^\dagger M_\nu)^2   M_\nu^*   \, h_e^*  M_\nu  \right] = 0~,  \\
\tilde{B}_3 & \equiv & {\rm Im} \, {\rm Tr} \ \left[   h_\nu  \,
(M_\nu^\dagger M_\nu)^2   M_\nu^*   \, h_e^*  M_\nu (M_\nu^\dagger M_\nu)
\right] = 0  \ . \label{eq:invariants2}
\end{eqnarray}

Note that  $B_{1,2,3}$ and $\tilde{B}_{1,2,3}$ are   
in direct correspondence with  (linearly independent combinations of) the CP
asymmetries $\epsilon_{\alpha}^{(j,i)}$ relevant for flavoured leptogenesis. In particular, 
$B_{1,2,3}$ correspond to the combinations %$\epsilon_{i}=\sum_\alpha \epsilon_{i \alpha}$ 
relevant for the 1-flavour regime.
This can be seen by working in the weak-basis where $M_\nu$ and $\lambda_e$ are diagonal.  
In this basis, for example, $B_1$  and $\tilde{B}_1$ read:

\begin{eqnarray}
B_1 &=&  \sum_{i<j} \,  M_i M_j (M_j^2 - M_i^2) 
 \sum_{\alpha=e,\mu,\tau} 
 {\rm Im} \left[  ({\lambda_\nu})_{i \alpha}   \,  ( {\lambda}_\nu^\dagger)_{\alpha j} \,   
 ( \lambda_\nu  \lambda_\nu^\dagger)_{ij}
 \right]
% \ {\rm Im} \left( (\bar
%\lambda_\nu \bar \lambda_\nu^\dagger)_{ij}^2 \right)
%
% +  M_1 M_3 (M_3^2 - M_1^2)
% \  {\rm Im} \left( (\bar \lambda_\nu \bar \lambda_\nu^\dagger)_{13}^2 \right)
%   \nonumber \\
%  &+&    M_2 M_3 (M_3^2 - M_2^2)
% \  {\rm Im} \left( (\bar \lambda_\nu \bar \lambda_\nu^\dagger)_{23}^2 \right)~.
   \nonumber \\
\tilde{B}_1 &=&  \sum_{i<j} \,   M_i M_j (M_j^2 - M_i^2)  
 \sum_{\alpha=e,\mu,\tau} 
\frac{m_\alpha^2}{v^2} \ 
 {\rm Im} \left[  ({\lambda_\nu})_{i \alpha}   \,  ( {\lambda}_\nu^\dagger)_{\alpha j} \,   
 ( \lambda_\nu  \lambda_\nu^\dagger)_{ij}
 \right]~.
\end{eqnarray}

%\footnote{
%In proving some of the properties of the weak-basis invariants, it is 
%useful to keep in mind the form that the various matrices  take in the "high scale" weak basis, {\it i.e.} 
%before diagonalising  the R-handed mass matrix.  This is the basis  denoted by 
%a "zero" superscript in Isabella's notes. In this basis  $\lambda_{e}$ is diagonal and 
%$h_\nu =  M_\nu/v^2   H  \hat{m}_\nu H$. }

If $M_\nu$ is proportional to the identity, the hermiticity of $h_{\nu,e}$ and the
cyclic property of the trace operation imply that the $B_{i}, \tilde{B}_i$ vanish identically.
The next step is to break the degeneracy of the heavy neutrinos in a way
consistent with the MLFV hypothesis. Selecting for convenience the 
0-superscript basis, let us now investigate under which conditions on $M_\nu$ of Eq.(\ref{mnu}) and 
$\lambda^0_\nu$ of Eq.(\ref{la0nu}) the $B_{i}, \tilde{B}_i$ are non-vanishing. 

\begin{itemize}

\item If we confine ourselves to terms quadratic in the Yukawa couplings, {\it i.e.} proportional to $c^{(1)}$, 
then we have that:
\begin{itemize}
\item  $B_{i}=0$ because of the hermiticity of $h^0_\nu$ and properties of the trace operator.
%
% ${\rm Im}  \, {\rm Tr} \, \left[ (h_\nu)^a (h_\nu^*)^b \right] =0$. 
%
This implies that in the unflavoured regime the CP asymmetries vanish;  

\item  $\tilde{B}_i  \neq 0$
%  \sim  {\rm Im}  \, {\rm Tr} \, \left[ (h_\nu)^a (h_\nu^*)^b h_e^* \right]  \neq 0$
as long as $H\neq I$. Since $\tilde{B}_i \propto c^{(1)} {\rm Im}  \, {\rm Tr} \, \left[ (h^0_\nu)^a  h_e^* \right] =  0$ ($a$ is some integer)  
and the restriction $H=I$ implies that $h^0_\nu$ is real, it turns out that leptogenesis is 
possible in the flavoured regime only if $H \neq I$.   
\end{itemize}

\item  If we include those terms  in $M_R$ that are  quartic in the Yukawa couplings, {\it i.e.}  
proportional to $c^{(2)}_{1,2,3,4}$, then:
\begin{itemize}
\item $B_i \neq 0$ if any of the $c^{(2)}_{i} \neq 0$, 
as long as $H \neq I$. 
If $H=I$ then not only $h^0_\nu$ is real, but also
$M_\nu$ and therefore $B_{1,2,3}=0$.
So one concludes that with $H=I$ unflavoured leptogenesis is not viable;

\item $\tilde{B}_i \neq 0$  if any of the $c^{(2)}_{i} \neq 0$. 
In this case, even setting $H=I$ %(real $h_\nu$ and $M_R$)
leads to  non-zero $\tilde{B}_i$  as long as $c^{(2)}_{4} \neq 0$.
So one concludes that leptogenesis with $H=I$ is 
potentially viable only in the flavoured regime with $c^{(2)}_{4} \neq 0$. 

\end{itemize} 
\end{itemize}

In conclusion, flavour effects open at least in principle two new regimes for 
MLFV-leptogenesis which are not allowed in the 1-flavour limit:  
(i) the case in which RH mass splitting is induced only (or mainly) by $c^{(1)}$. This situation requires $H\neq I$,
namely CPV in the RH sector;
(ii) the case in which CPV arises only from PMNS phases, namely $H=I$ and $M_\nu$ is real.

%%%%%%%%%%%%%%%%%%%%%%%%%%%%%%%%%%%%%%%%%%%%%%%%%%%%%


\begin{thebibliography}{99}

\bibitem{wmap}  D.~N.~Spergel {\it et al.}  [WMAP Collaboration],
 arXiv:astro-ph/0603449. 

\bibitem{FY} M.~Fukugita and T.~Yanagida,
  %``Baryogenesis Without Grand Unification,''
  Phys.\ Lett.\ B {\bf 174}, 45 (1986).

\bibitem{leptogen} G.~F.~Giudice, A.~Notari, M.~Raidal, A.~Riotto and A.~Strumia,
%``Towards a complete theory of thermal leptogenesis in the SM and MSSM,''
Nucl.\ Phys.\  B {\bf 685}, 89 (2004) [arXiv:hep-ph/0310123];
%%CITATION = HEP-PH 0310123;%%
W.~Buchmuller, P.~Di Bari and M.~Plumacher,
%``Leptogenesis for pedestrians,''
Annals Phys.\  {\bf 315} (2005) 305 [arXiv:hep-ph/0401240].
%%CITATION = HEP-PH 0401240;%%

\bibitem{work} A partial list:~W.~Buchmuller, P.~Di Bari and M.~Plumacher,
Nucl.\ Phys.\ B {\bf 643} (2002) 367 [arXiv:hep-ph/0205349];
%%CITATION = HEP-PH 0205349;%%
J.~R.~Ellis, M.~Raidal and T.~Yanagida,
%``Observable consequences of partially degenerate leptogenesis,''
Phys.\ Lett.\ B {\bf 546} (2002) 228 [arXiv:hep-ph/0206300];
%%CITATION = HEP-PH 0206300;%%
G.~C.~Branco, R.~Gonzalez Felipe, F.~R.~Joaquim and M.~N.~Rebelo,
%``Leptogenesis, CP violation and neutrino data: What can we learn?,''
Nucl.\ Phys.\  B {\bf 640} (2002) 202 [arXiv:hep-ph/0202030];
%%CITATION = HEP-PH 0202030;%%
G.~C.~Branco, R.~Gonzalez Felipe, F.~R.~Joaquim, I.~Masina,
M.~N.~Rebelo and C.~A.~Savoy,
%``Minimal scenarios for leptogenesis and CP violation,''
Phys.\ Rev.\ D {\bf 67}, 073025 (2003) [arXiv:hep-ph/0211001];
%%CITATION = HEP-PH 0211001;%%
R.~N.~Mohapatra, S.~Nasri and H.~B.~Yu,
%``Leptogenesis, mu - tau symmetry and theta(13),''
Phys.\ Lett.\  B {\bf 615} (2005) 231 [arXiv:hep-ph/0502026];
%%CITATION = HEP-PH 0502026;%%
A.~Broncano, M.~B.~Gavela and E.~Jenkins,
%``Neutrino physics in the seesaw model,''
Nucl.\ Phys.\  B {\bf 672} (2003) 163 [arXiv:hep-ph/0307058];
%%CITATION = HEP-PH 0307058;%%
A.~Pilaftsis,
%``CP violation and baryogenesis due to heavy Majorana neutrinos,''
Phys.\ Rev.\ D {\bf 56} (1997) 5431 [arXiv:hep-ph/9707235];
%%CITATION = HEP-PH 9707235;%%
E.~Nezri and J.~Orloff,
%``Neutrino oscillations vs. leptogenesis in S{\cal O}(10) models,''
JHEP {\bf 0304} (2003) 020 [arXiv:hep-ph/0004227];
%%CITATION = HEP-PH 0004227;%%
S.~Davidson and A.~Ibarra,
%``Leptogenesis and low-energy phases,''
Nucl.\ Phys.\ B {\bf 648}, 345 (2003) [arXiv:hep-ph/0206304];
%%CITATION = HEP-PH 0206304;%%
S.~Davidson,
%``From weak-scale observables to leptogenesis,''
JHEP {\bf 0303} (2003) 037 [arXiv:hep-ph/0302075];
%%CITATION = HEP-PH 0302075;%%
S.~T.~Petcov, W.~Rodejohann, T.~Shindou and Y.~Takanishi,
%``The see-saw mechanism, neutrino Yukawa couplings, LFV decays l(i) ...
Nucl.\ Phys.\ B {\bf 739} (2006) 208 [arXiv:hep-ph/0510404].
%%CITATION = HEP-PH 0510404;%%



\bibitem{seesaw} 
P.~Minkowski,
%``Mu $\to$ E Gamma At A Rate Of One Out Of 1-Billion Muon Decays?,''
Phys.\ Lett.\ B {\bf 67} (1977) 421;
%%CITATION = PHLTA,B67,421;%%
M. Gell-Mann, P. Ramond and
R. Slansky,  {\em Proceedings of the Supergravity Stony Brook Workshop}, New
York 1979,  eds. P. Van Nieuwenhuizen and D. Freedman; T. Yanagida,  {\em
Proceedinds of the Workshop on Unified Theories and Baryon Number in the
Universe},  Tsukuba, Japan 1979, ed.s A. Sawada and A. Sugamoto;
R. N. Mohapatra, G. Senjanovic,
{\it Phys.Rev.Lett.} {\bf 44} (1980)912.

\bibitem{sakharov} A.D. Sakharov.   JETP Lett. {\bf 5} (1967) 24.

\bibitem{baureview} 
For a review, see A.~Riotto and M.~Trodden,
  %``Recent progress in baryogenesis,''
  Ann.\ Rev.\ Nucl.\ Part.\ Sci.\  {\bf 49}, 35 (1999).

\bibitem{kuzmin} V.A. Kuzmin, V.A. Rubakov, and M.E. Shaposhnikov. \newblock
{\em Phys. Lett.}, B155:36, 1985.


\bibitem{di} S.~Davidson and A.~Ibarra,
  %``A lower bound on the right-handed neutrino mass from leptogenesis,''
  Phys.\ Lett.\  B {\bf 535}, 25 (2002). 



\bibitem{davidsonetal1} 
A.~Abada, S.~Davidson, F.~X.~Josse-Michaux, M.~Losada and A.~Riotto,
  %``Flavor issues in leptogenesis,''
  JCAP {\bf 0604}, 004 (2006).

\bibitem{aa} S.~Antusch and A.~M.~Teixeira,
  %``Towards constraints on the SUSY seesaw from flavor-dependent
  %leptogenesis,''
  JCAP {\bf 0702}, 024 (2007).

\bibitem{jmr} F.~R.~Joaquim, I.~Masina and A.~Riotto,
  %``Observable electron EDM and leptogenesis,''
  arXiv:hep-ph/0701270.


\bibitem{grav} For a recent review, see
T.~Moroi,
%``Gravitino production in the early universe and its implications to
AIP Conf.\ Proc.\  {\bf 805}, 37 (2006).
%%CITATION = HEP-PH 0509121;%%

\bibitem{res} 
M.~Flanz, E.~A.~Paschos and U.~Sarkar,
%``Baryogenesis from a lepton asymmetric universe,''
Phys.\ Lett.\ B {\bf 345} (1995) 248 [Erratum-ibid.\ B {\bf 382}
(1996) 447] [arXiv:hep-ph/9411366];
L.~Covi and E.~Roulet,
  %``Baryogenesis from mixed particle decays,''
  Phys.\ Lett.\  B {\bf 399}, 113 (1997) [arXiv:hep-ph/9611425];
A.~Pilaftsis,
%``CP violation and baryogenesis due to heavy Majorana neutrinos,''
Phys.\ Rev.\ D {\bf 56}, 5431 (1997) [arXiv:hep-ph/9707235];
T.~Hambye,
  %``Leptogenesis at the TeV scale,''
  Nucl.\ Phys.\  B {\bf 633}, 171 (2002) [arXiv:hep-ph/0111089]; 
A.~Pilaftsis and T.~E.~J.~Underwood,
  %``Resonant leptogenesis,''
  Nucl.\ Phys.\  B {\bf 692}, 303 (2004) [arXiv:hep-ph/0309342];
A.~Pilaftsis and T.~E.~J.~Underwood,
  %``Electroweak-scale resonant leptogenesis,''
  Phys.\ Rev.\  D {\bf 72}, 113001 (2005) [arXiv:hep-ph/0506107].



\bibitem{MQFV}
%\bibitem{Chivukula:1987py}
  R.~S.~Chivukula and H.~Georgi,
  %``Composite Technicolor Standard Model,''
  Phys.\ Lett.\  B {\bf 188} (1987) 99; 
  %%CITATION = PHLTA,B188,99;%%
%\bibitem{Hall:1990ac}
  L.~J.~Hall and L.~Randall,
  %``Weak scale effective supersymmetry,''
  Phys.\ Rev.\ Lett.\  {\bf 65} (1990) 2939.
  %%CITATION = PRLTA,65,2939;%%

\bibitem{MFV}
  G.~D'Ambrosio, G.~F.~Giudice, G.~Isidori and A.~Strumia,
  %``Minimal flavor violation: An effective field theory approach,''
  Nucl.\ Phys.\  B {\bf 645} (2002) 155
  [arXiv:hep-ph/0207036].
  %%CITATION = NUPHA,B645,155;%%


\bibitem{Cirigliano:2005ck}
  V.~Cirigliano, B.~Grinstein, G.~Isidori and M.~B.~Wise,
  %``Minimal flavor violation in the lepton sector,''
  Nucl.\ Phys.\  B {\bf 728} (2005) 121
  [arXiv:hep-ph/0507001].
  %%CITATION = NUPHA,B728,121;%%


\bibitem{Cirigliano:2006su}
  V.~Cirigliano and B.~Grinstein,
  %``Phenomenology of minimal lepton flavor violation,''
  Nucl.\ Phys.\  B {\bf 752} (2006) 18
  [arXiv:hep-ph/0601111].
  %%CITATION = NUPHA,B752,18;%%


\bibitem{Grinstein:2006cg}
  B.~Grinstein, V.~Cirigliano, G.~Isidori and M.~B.~Wise,
  %``Grand unification and the principle of minimal flavor violation,''
  Nucl.\ Phys.\  B {\bf 763} (2007) 35
  [arXiv:hep-ph/0608123].
  %%CITATION = NUPHA,B763,35;%%

\bibitem{Davidson:2006bd}
  S.~Davidson and F.~Palorini,
  %``Various definitions of minimal flavor violation for leptons,''
  Phys.\ Lett.\  B {\bf 642} (2006) 72.

\bibitem{Cirigliano:2006nu}
  V.~Cirigliano, G.~Isidori and V.~Porretti,
  %``CP violation and leptogenesis in models with minimal lepton flavor
  %violation,''
  Nucl.\ Phys.\  B {\bf 763} (2007) 228
  [arXiv:hep-ph/0607068].
  %%CITATION = NUPHA,B763,228;%%



\bibitem{Branco:2006hz}
  G.~C.~Branco, A.~J.~Buras, S.~Jager, S.~Uhlig and A.~Weiler,
  %``Another look at minimal lepton flavor violation, l(i) --> l(j) gamma,
  %leptogenesis, and the ratio M(nu)/Lambda(LFV),''
  JHEP {\bf 0709} (2007) 004
  [arXiv:hep-ph/0609067].
  %%CITATION = JHEPA,0709,004;%%


\bibitem{Uhlig:2006xf}
  S.~Uhlig,
  %``Minimal lepton flavor violation and leptogenesis with exclusively
  %low-energy CP violation,''
  arXiv:hep-ph/0612262.
  %%CITATION = HEP-PH/0612262;%%








\bibitem{dsr1}
  A.~De Simone and A.~Riotto,
  %``Quantum Boltzmann equations and leptogenesis,''
  JCAP {\bf 0708} (2007) 002
  [arXiv:hep-ph/0703175].
  %%CITATION = JCAPA,0708,002;%%

\bibitem{buchmuller}  W.~Buchmuller and S.~Fredenhagen,
  %``Quantum mechanics of baryogenesis,''
  Phys.\ Lett.\  B {\bf 483}, 217 (2000) [arXiv:hep-ph/0004145].
  
\bibitem{dsr2}
  A.~De Simone and A.~Riotto,
  %``On resonant leptogenesis,''
  JCAP {\bf 0708} (2007) 013
  [arXiv:0705.2183 [hep-ph].
  %%CITATION = JCAPA,0708,013;%%

\bibitem{f1} R.~Barbieri, P.~Creminelli, A.~Strumia and N.~Tetradis,
  %``Baryogenesis through leptogenesis,''
  Nucl.\ Phys.\  B {\bf 575} (2000) 61; T.~Endoh, T.~Morozumi and Z.~h.~Xiong,
  %``Primordial lepton family asymmetries in seesaw model,''
  Prog.\ Theor.\ Phys.\  {\bf 111} (2004) 123;

\bibitem{f2} A.~Abada, S.~Davidson, F.~X.~Josse-Michaux, M.~Losada and A.~Riotto,
  %``Flavor issues in leptogenesis,''
  JCAP {\bf 0604}, 004 (2006); 
E.~Nardi, Y.~Nir, E.~Roulet and J.~Racker,
%``The importance of flavor in leptogenesis,''
JHEP {\bf 0601}, 164 (2006).

\bibitem{f3} A.~Abada, S.~Davidson, A.~Ibarra, F.~X.~Josse-Michaux, M.~Losada and A.~Riotto,
  %``Flavor matters in leptogenesis,''
  JHEP {\bf 0609}, 010 (2006).

\bibitem{f4} S.~Blanchet and P.~Di Bari,
  %``Flavor effects on leptogenesis predictions,''
  JCAP {\bf 0703}, 018 (2007);
S.~Antusch, S.~F.~King and A.~Riotto,
%``Flavor-dependent leptogenesis with sequential dominance,''
JCAP {\bf 0611}, 011 (2006);  S.~Pascoli, S.~T.~Petcov and A.~Riotto, 
.\ Rev.\  D {\bf 75}, 083511 (2007); 
 G.~C.~Branco, R.~Gonzalez Felipe and F.~R.~Joaquim,
  %``A new bridge between leptonic CP violation and leptogenesis,''
  Phys.\ Lett.\  B {\bf 645} (2007) 432; 
 S.~Pascoli, S.~T.~Petcov and A.~Riotto;
.\ Rev.\  D {\bf 75}, 083511 (2007); 
 G.~Engelhard, Y.~Grossman, E.~Nardi and Y.~Nir,
hep-ph/0612187;
S.~Blanchet, P.~Di Bari and G.~G.~Raffelt,
  %``Quantum Zeno effect and the impact of flavor in leptogenesis,''
  JCAP {\bf 0703}, 012 (2007);
S.~Pascoli, S.~T.~Petcov and A.~Riotto,
  %``Leptogenesis and low energy CP violation in neutrino physics,''
  Nucl.\ Phys.\  B {\bf 774}, 1 (2007); 
A.~De Simone and A.~Riotto,
  %``On the impact of flavor oscillations in leptogenesis,''
  JCAP {\bf 0702} (2007) 005; 
T.~Shindou and T.~Yamashita,
  %``A novel washout effect in the flavored leptogenesis,''
 JHEP {\bf 0709}, 043 (2007); 
F.~X.~Josse-Michaux and A.~Abada,
  %``Study of flavor dependencies in leptogenesis,''
  arXiv:hep-ph/0703084. 














\bibitem{Casas:2001sr}
J.~A.~Casas and A.~Ibarra,
%``Oscillating neutrinos and mu $\to$ e, gamma,''
Nucl.\ Phys.\ B {\bf 618} (2001) 171
[arXiv:hep-ph/0103065].
%%CITATION = HEP-PH 0103065;%%


\bibitem{Branco:2001pq}
  G.~C.~Branco, T.~Morozumi, B.~M.~Nobre and M.~N.~Rebelo,
  %``A bridge between CP violation at low energies and leptogenesis,''
  Nucl.\ Phys.\ B {\bf 617}, 475 (2001)
  [arXiv:hep-ph/0107164].
  %%CITATION = HEP-PH 0107164;%%



\end{thebibliography}
\end{document}